\documentclass[10pt,a4paper]{article}
\usepackage[dvips]{color}
\usepackage{epsfig}
\usepackage{epstopdf}
\usepackage{amsmath}
\usepackage{graphicx}
\usepackage{authblk}
\usepackage{amssymb,amsmath}
\usepackage{amsmath}
\usepackage{tabularx}

\begin{document}
\title{\bf Non-linear logarithmic interactions and a varying polytropic gas}
\author{{M. Khurshudyan$^{a,b,c}$\thanks{Email:khurshudyan@yandex.ru, khurshudyan@tusur.ru}~~, M. Hakobyan$^{c}$\thanks{Email:hakobyan1margarit@gmail.com}~~, As. Khurshudyan$^{d}$\thanks{Email:khurshudyan@mechins.sci.am}}\\
$^{a}${\small {\em International Laboratory for Theoretical Cosmology, TUSUR, Tomsk, Russia}}\\ 
$^{b}${\small {\em Research Division, TSPU, Tomsk, Russia}}\\
$^{c}${\small {\em Institute of Physics, University of Zielona Gora, Zielona Gora, Poland}}\\
$^{d}${\small {\em Institute of Mechanics, National Academy of Sciences of Armenia, Yerevan, Armenia}}\\
}\maketitle

\begin{abstract}
In this paper we will demonstrate possible existence of more exotic forms of interaction taking into account, that dark energy can be parametrized as a varying polytropic fluid suggested recently by the first author. On going research in both directions, i.e. dark energy and non-gravitational interaction, opens very interesting perspectives for solving the problems which are related to the accelerated expansion of the Universe. The fact that available observational data allows to parametrize the dark side of our Universe in form of interacting dark fluids is not less surprising, than the accelerated expansion itself. Therefore, the consideration of new forms of non-gravitational interactions in modern cosmology still is one of the actual topics and requires deep and systematic research. This fact motivates us to consider various new forms of non-gravitational logarithmic interactions and study their cosmological consequences. Appropriate classification of the models is presented in the light of actively discussed $Om$ analysis and the result for the Hubble parameter measured at $z = 2.34$ in BOSS experiment. 

\end{abstract}

\section{Introduction}\label{sec:INT}
A continuous research towards understanding of a particular regime in the evolution of the Universe defined for the redshifts up to $z\approx 3$ uses various phenomenological assumptions. In particular, when the background dynamics of the Universe is assumed to be according to general relativity, two main phenomenological assumptions are about an existence of a certain amount of dark energy and dark matter allowing the evolution of our observable Universe. On the other hand, it is well know that the Universe with $z < 3$ has a transition from the phase with decelerated expanding to an accelerated expanding phase, which is directly related to the physics of the early Universe. In case of early Universe, if we believe that general relativity can be applied, then we need to have inflation, which with appropriate lifetime provides all seeds for observable Universe. Moreover, for $z < 1$, the observational data besides to the fact that the recent Universe is in the accelerated expanding phase, does not allow to decrypt the structure of dark energy and dark matter~\cite{MR}. This brings to different ideas including to a modification of general relativity, which efficiently can be applied to understand the physics of the early Universe as well. One of the famous modifications of general relativity among the others is $f(R)$ theory. There is an active research towards modified theories of general relativity revealing its cosmological and astrophysical applications~\cite{MG1}~-~\cite{MG12} to mention a few. In this paper we will assume that general relativity can be applied to describe the background dynamics of the large scale Universe and to explain the accelerated expansion by hand we will include a particular model of dark fluid and non-gravitational interactions between it and cold dark matter with $P=0$. The dark energy model considered in this paper is a particular model of a varying polytropic dark fluid considered in Ref.~\cite{MK1}. Palytropic fluid itself represents a huge interest in astrophysics due to its nonlinear form of the equation of state. Recently it has been demonstrated that the parametrization of the dark side of the Universe in form of interacting varying polytropic fluid and cold dark matter can explain the accelerated expansion of the Universe in an efficient way. Moreover, specific forms of non-gravitational interaction allowed to obtain the solution of the cosmological coincidence problem~(see details in \cite{MK1}). This motivates us to consider new forms of non-gravitational interactions in considered parameterization allowing to extend existing study. The phenomenological form of the varying polytropic fluid presented in Ref.~\cite{MK1} has been constructed taking into account a phenomenological modification of Chaplygin gas. An interest towards Chaplygin gas, which also has nonlinear form of the equation of state, is due to the fact that it is a joint model of dark energy and dark matter for appropriate values of the parameters~\cite{MK2}~-~\cite{MK6}~(and references therein). In general, the phenomenology is actively involved in modern cosmology allowing to improve our knowledge about the physics of the Universe. Actively developed various phenomenological ideas, like for instance non-gravitational interaction, allows to improve theoretical results. One of the first examples presented in literature is related to the results obtained with holographic dark energy models, when it has been demonstrated that an existing interaction can explain the accelerated expansion and solve related problems relevant to such models. While in case of non interacting models it is not possible to do~(see for instance~\cite{MK2}~-~\cite{MK6} and references therein). In general, existing tension between different observational datasets creates a huge gap in theory and the phenomenology plays a crucial role to cover it. However, it is believed that updated data eventually will remove the tension and it will be possible to constrain the models in a very efficient way. However, we should remember that so far it is possible to scan the Universe up to redshifts $z \approx 3$ and for more constraints we should be able to scan the Universe with $z > 3$. One of such missions is BOSS experiment allowing, for instance, to estimate the value of the Hubble parameter for $z=2.34$~\cite{BOSS}. This estimation is unique by its nature and has been used in literature in order to constrain different new cosmological models. In particular in Ref.~\cite{BOSS1} the authors considered cosmological models involving dark energy with a constant equation of state parameter $\omega_{de} = P/\rho$ claimed that the value of the Hubble parameter at $z=2.34$ estimated from BOSS experiment can be a direct prove about an existence of non-gravitational interaction between dark energy and dark matter. On the other hand, they suspect that to explain the value of the Hubble parameter estimated from the same experiment will require some unusual form of dynamical dark energy. However, the authors did not mention how the form of the dynamical dark energy should be exotic. It is well known that dynamical dark energy models involved in modern cosmology can solve the problems associated to the $\Lambda$ model of dark energy~\cite{DE1}. In this paper, we will demonstrate when a dynamical dark energy model in form of polytropic dark fluid
\begin{equation}\label{eq:VP}
P_{de} = A H^{-k} \rho_{de}^{1 + 1/n}, 
\end{equation}            
introduced in Ref.~\cite{MK1} can explain mentioned BOSS experiment result for the Hubble parameter at $z=2.34$. In this study we used constraints originated form $Om$ analysis for mentioned purposes~\cite{SahniFin}.\\\\
The paper is organized as follows: in section~\ref{MPSA} we will discuss the models and the results from performed studies. This includes a detailed description of the background dynamics in form of cosmography, the forms of non-gravitational interactions and $Om$ analysis. In section~\ref{sec:Dis} we have organized discussion on obtained results and present possible extensions of the models considered in this work as a subject for next papers.

\section{Models}\label{MPSA}
For considered models of this paper we accepted general relativity to be the theory describing the background dynamics. Moreover, knowing in advance about geometry and symmetries of the large scale Universe, which can be described via FRW metric, with $8 \pi G = c = 1$ we assume that the effective fluid in the Universe can be presented in the following way
\begin{equation}\label{eq:rhoeff}
\rho_{eff} = \rho_{de} + \rho_{dm},
\end{equation} 
\begin{equation}\label{eq:Peff}
P_{eff} = P_{de},
\end{equation}
where $\rho_{de}$ and $\rho_{dm}$ are the energy densities of dark energy and dark matter, while $P_{de}$ is the pressures of the dark energy (the preasure of dark matter is $P_{dm} = 0 $). With this assumption, the dynamics of the energy densities will take the following form
\begin{equation}\label{eq:rhoDE}
\dot{\rho}_{de} + 3 H (\rho_{de} + P_{de}) = -Q,
\end{equation} 
\begin{equation}\label{eq:rhoDM}
\dot{\rho}_{dm} + 3 H \rho_{dm} = Q,
\end{equation} 
where $Q$ stands for the interaction inside the darkness, while the energy density of the effective fluid, Eq.~(\ref{eq:rhoeff}), allows to determine the Hubble parameter $H$ to be 
\begin{equation}\label{eq:Hubble}
H^{2} = \frac{1}{3} \rho_{eff}.
\end{equation}
Eq.~(\ref{eq:rhoDE}) and Eq.~(\ref{eq:rhoDM}) represent the idea of the non-gravitational interaction between dark energy and dark matter introduced in general relativity for the accelerated expanding Universe. The recent study indicates that dark energy non-gravitationally is coupled to dark matter, rather than to other existing species  in Universe~(due to recent observational data). There are various geometrical tools in order to study dark energy and cosmological models and one of them is $Om$ analysis
\begin{equation}
Om = \frac{x^{2}-1}{(1+z)^{3} - 1},
\end{equation}  
where $x = H/H_{0}$ and $H_{0}$ to be the value of the Hubble parameter at $z=0$. It is assumed that if different trajectories have been obtained, then the models are different. In this paper we will use a small modification of two point $Om$~($Omh^{2}$) analysis and known results from it for $z_{1} = 0$, $z_{2} = 0.57$ and $z_{3} = 2.34$~\cite{SahniFin}
$$Omh^{2}(z_{1};z_{2}) = 0.124 \pm 0.045,$$
$$Omh^{2}(z_{1};z_{3}) = 0.122 \pm 0.01,$$
\begin{equation}
Omh^{2}(z_{2};z_{3}) = 0.122 \pm 0.012,
\end{equation}
where 
\begin{equation}
Om(z_{2},z_{1}) = \frac{x(z_{2})^{2} - x(z_{1}^{2})}{(1+z_{2})^{2} - (1+z_{1})^{2}},
\end{equation}
while for $\Lambda$CDM the value is $Omh^{2} = 0.1426$, in order to constrain the considered models. We start the analysis from non-interacting model in next subsection.

\subsection{Non interacting case} 

\begin{figure}[h!]
 \begin{center}$
 \begin{array}{cccc}
\includegraphics[width=80 mm]{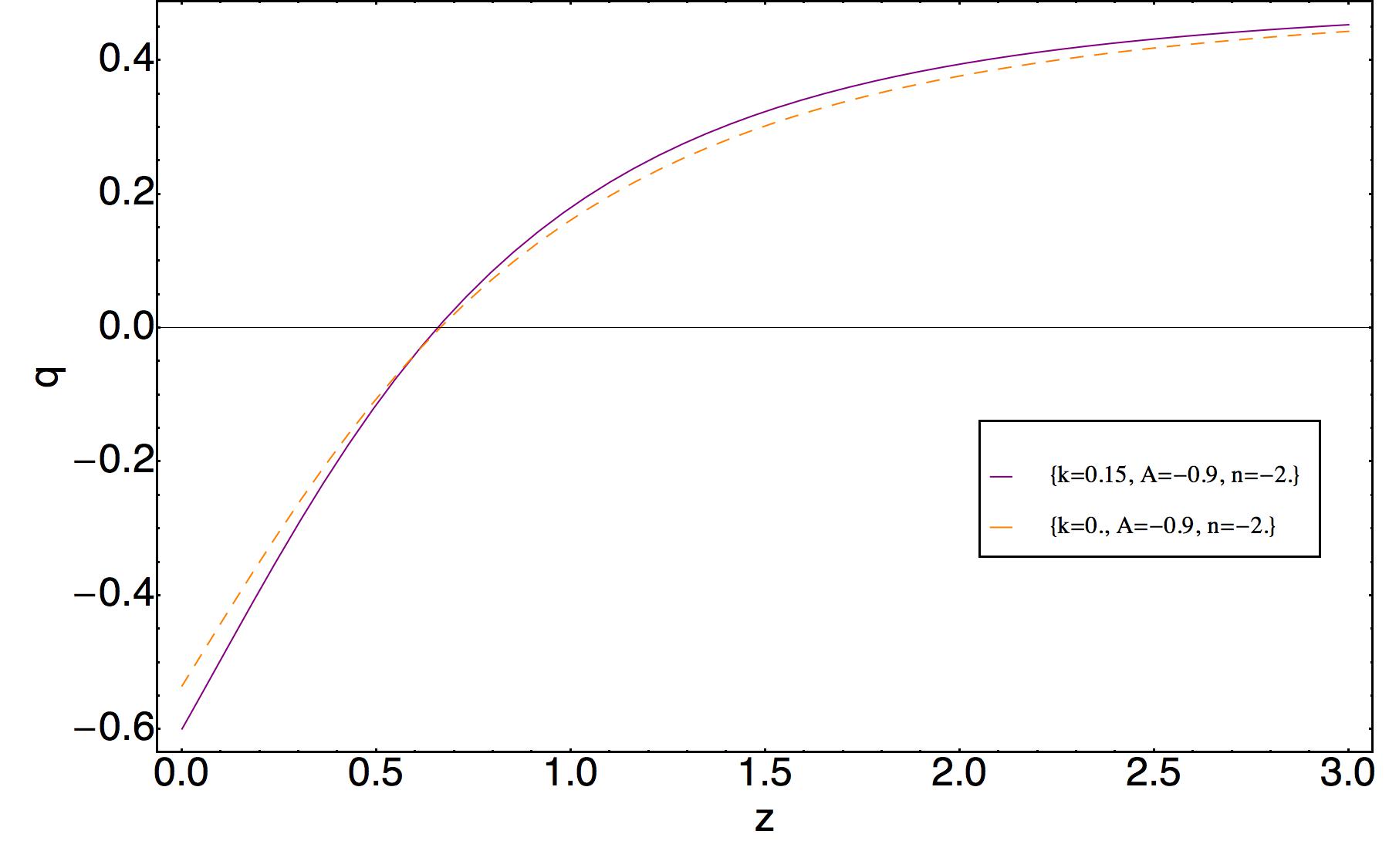} &
\includegraphics[width=80 mm]{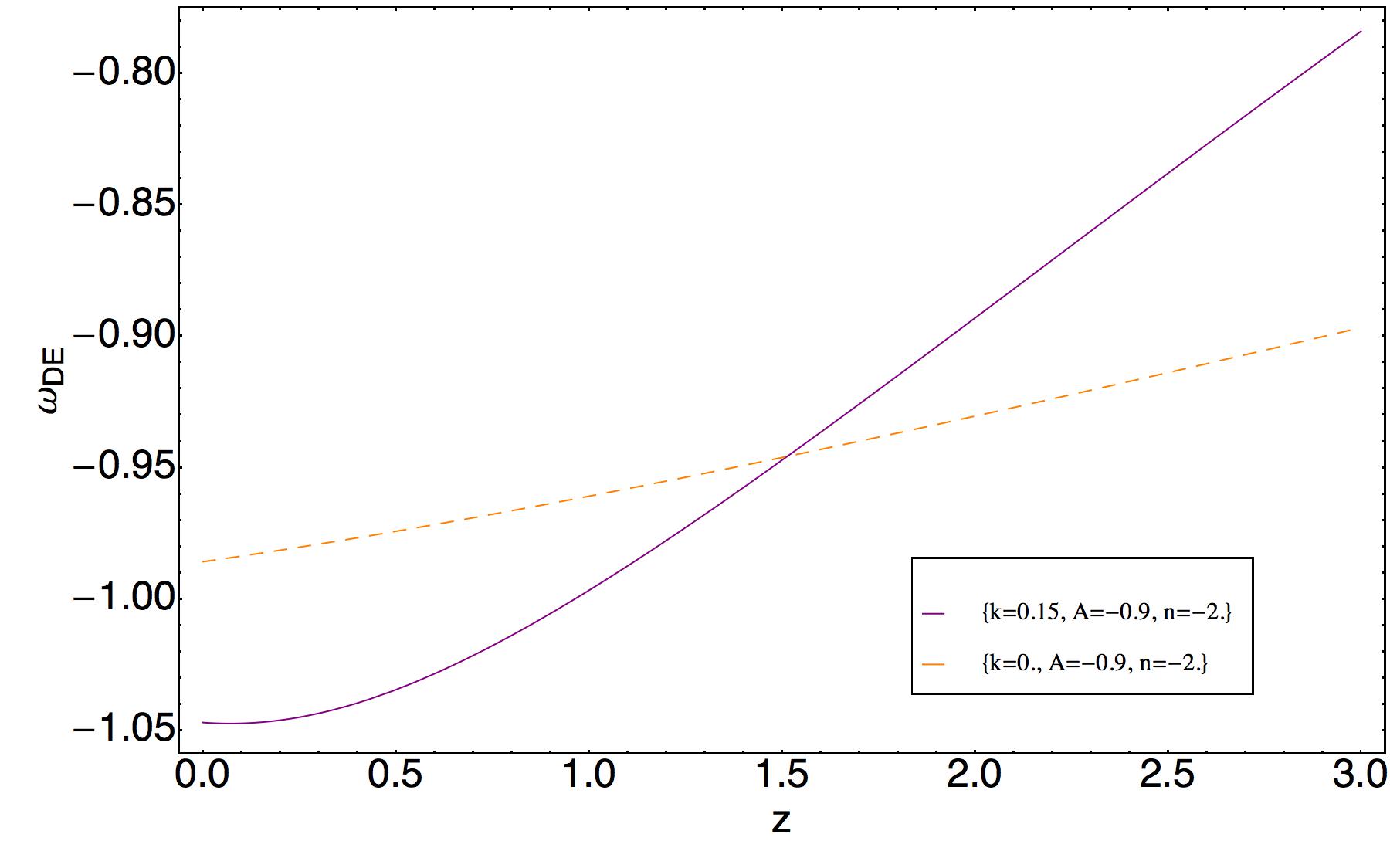}  
 \end{array}$
 \end{center}
\caption{The graphical behavior of the deceleration parameter $q$ and the equation of state parameter $\omega_{de}$ against the redshift $z$. The considered non-interacting models of polytropic and varying polytropic dark fluids provide cosmological models free from the cosmological coincidence problem.}
 \label{fig:Fig1}
\end{figure}

In general, non-interacting models by their simple structure can be integrated analytically. However, in this paper we will study the model numerically demonstrating the main features via graphical behavior of the parameters. The graphical behavior of the deceleration parameter $q$ and the equation of state~(EoS) parameter $\omega_{de}$ presented in Fig.~(\ref{fig:Fig1}) represents the comparison between the cosmological models containing non-interacting polytropic and non-interacting varying polytropic fluids. The EoS of the varying polytropic fluid is given by Eq.~(\ref{eq:VP}). The comparison of the graphical behavior of the deceleration parameter for both models shows that some differences can be observed for $z > 0.7$ and $z < 0.5$, respectively. Moreover, for both models the transition redshift is the same i.e. the parameter $k$ raised due to the modification does not play a crucial role on the transition redshift. However, the comparison of the $\omega_{de}$ presented on the right plot of Fig.~(\ref{fig:Fig1}) shows that the parameter $k$ significantly affects on the behavior of this parameters indicating differences between the models. In particular, EoS parameter for the non-interacting polytropic dark fluid~($k=0$) is linearly decreasing function of the redshift $z$ and for $z \in [0,3]$ the fluid has only quintessence nature. On the other hand, the equation of state parameter for the non-interacting varying polytropic dark fluid is a linearly decreasing function up to $z \approx 1.1$, then being again a decreasing function started from $z \approx 0.3$ becomes a constant. On the other hand, in this case the phantom line crossing is possible, since $\omega_{de} \approx - 1.05$. Moreover, the present day value of the EoS parameter is within the constraints obtained from PLANCK 2015 experiments~\cite{PC}. 

\begin{figure}[h!]
 \begin{center}$
 \begin{array}{cccc}
\includegraphics[width=80 mm]{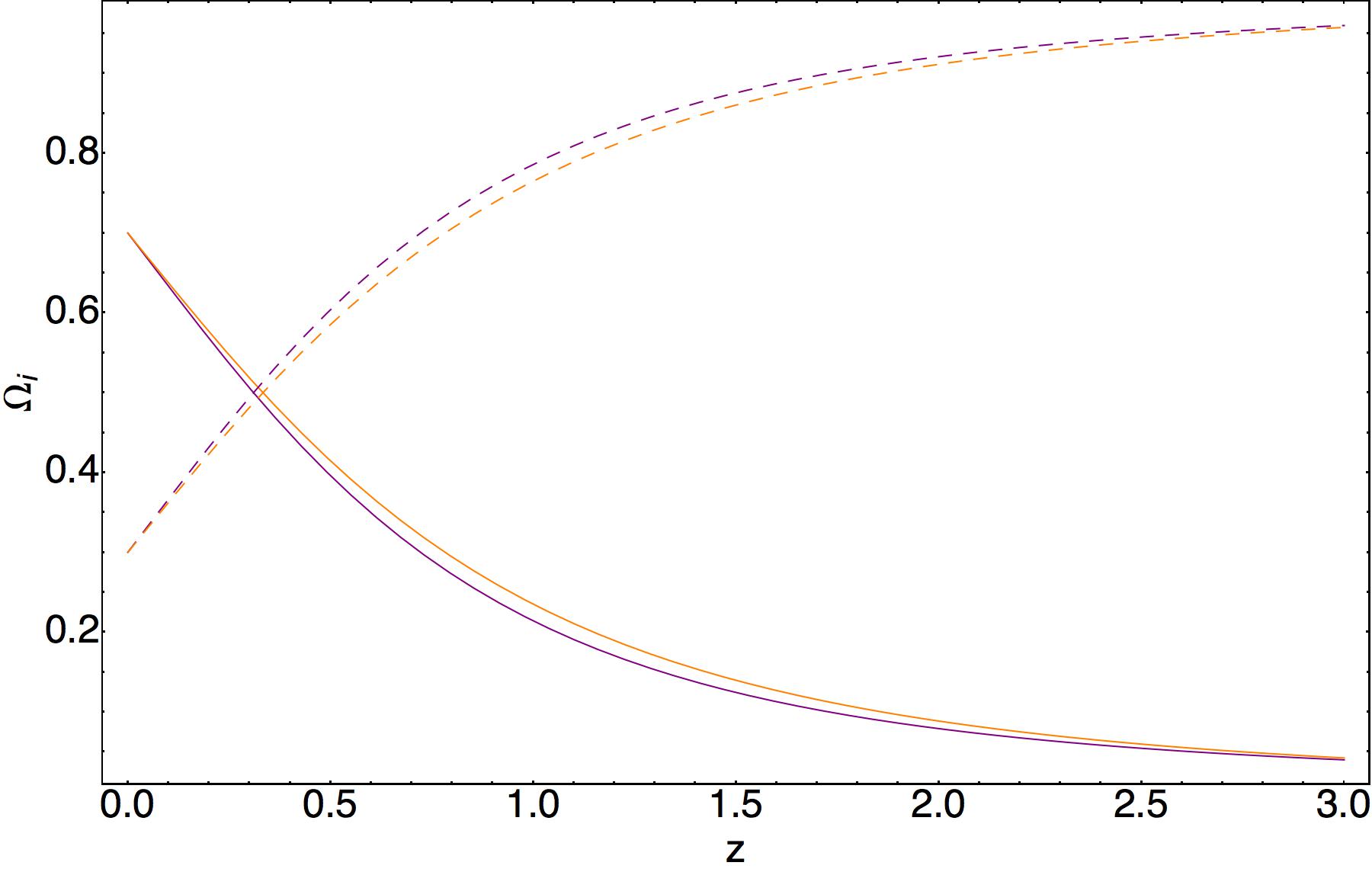} &
\includegraphics[width=80 mm]{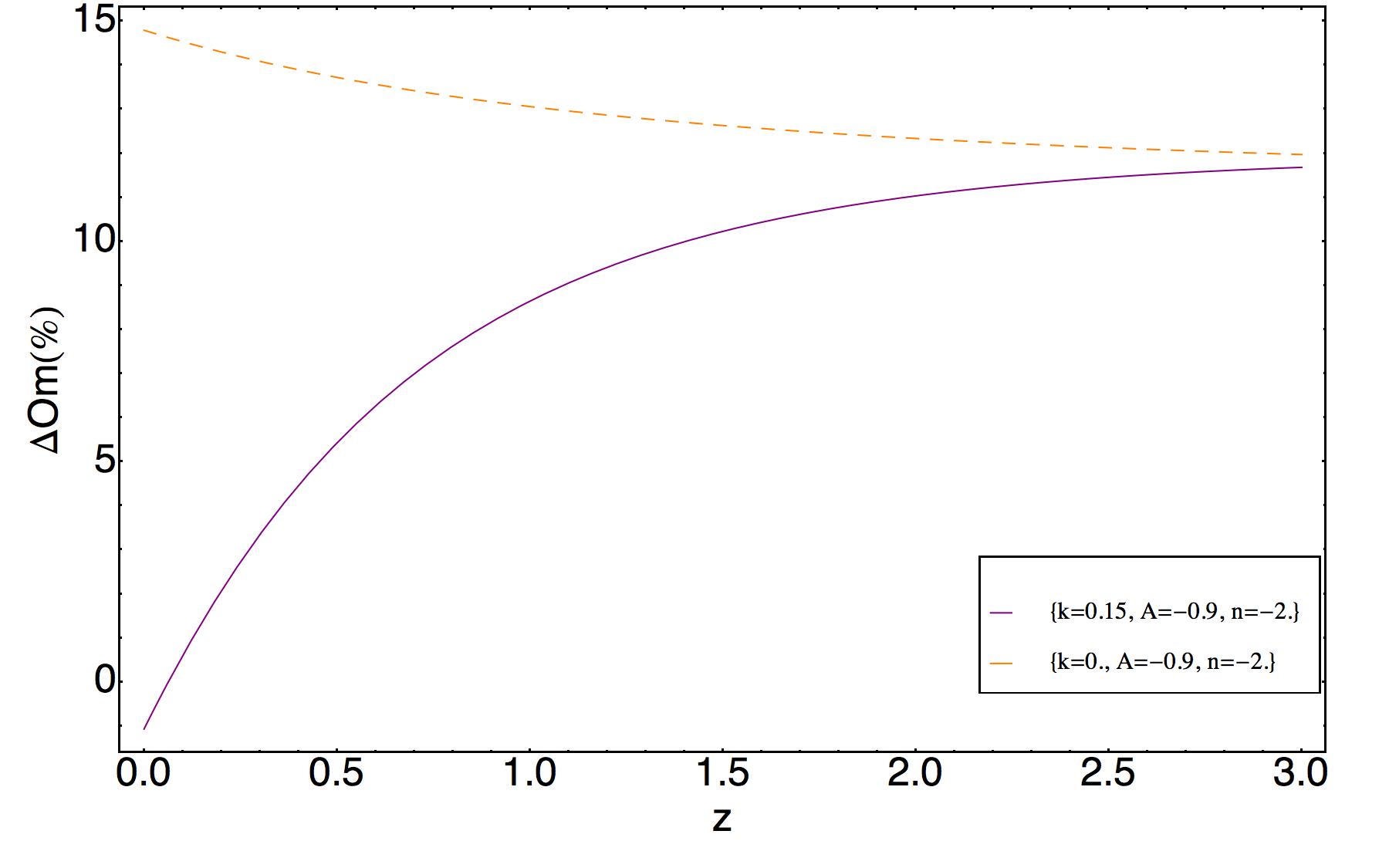}  
 \end{array}$
 \end{center}
\caption{The graphical behavior of $\Omega_{de}$ and $\Omega_{dm}$ against the redshift $z$ is presented on the left plot, while the right plot represents $\Delta Om$ for non-interacting polytropic and varying polytropic dark fluids providing cosmological models free from the cosmological coincidence problem. Presented behavior for $\Omega_{de}$ and $\Omega_{dm}$ is according to the same values of the parameters as for the behavior of the deceleration parameter $q$. The solid lines on $\Omega_{i} - z$ plane represent the behavior of $\Omega_{de}$, while dashed lines represent the behavior of $\Omega_{dm}$.}
 \label{fig:Fig2}
\end{figure}

The graphical behavior of $\Omega_{de}$ and $\Omega_{dm}$ parameters presented in Fig.~(\ref{fig:Fig2}) indicates that the models are free from the cosmological coincidence problem and that for higher and lower redshifts the behavior of these parameters with a high precision is the same. In other words, the deceleration parameter with $\Omega_{de}$ and $\Omega_{dm}$ parameters are not enough for deep and conclusive study of the models. Completely another picture have been observed when the models have been studied via $Om$ analysis. In particular, the right plot of Fig.~(\ref{fig:Fig2}) represents the graphical behavior of $\Delta Om$
\begin{equation}\label{eq:DeltaOm}
\Delta Om ( \% )= 100 \times \left [ \frac{Om_{model}}{Om_{\Lambda CDM} } - 1 \right ],
\end{equation}
where $Om_{\Lambda CDM} = \Omega^{(0)}_{dm} = 0.27$, allowing to estimate differences between $\Lambda$CDM standard model and suggested new models. Moreover, the study allows to demonstrate differences between new models as it can be seen from  Fig.~(\ref{fig:Fig2}). From the graphical behavior of the blue curve representing the cosmological model with non-interacting varying polytropic fluid it can be seen that for higher redshifts the relative difference is about $12\%$. However, the difference will decrease and at $z \approx 0.1$ it will disappear. On the other hand, for the redshifts $z < 0.1$ the relative difference is about $1\%$. Moreover, the relative difference between the $\Lambda$CDM standard model and the model with polytropic dark fluid will only will increase and at $z=0$ it is about $15\%$. For both models studied in this subsection we used the values of the parameters obtained from the constraints imposed by $Omh^{2}$ discussed at the beginning of this section~(Table~\ref{tab:Table1}). In particular, we see that the estimated value of the Hubble parameter at $z=2.34$ for both models are in well correspondence with the results from BOSS experiment, while the values of the Hubble parameters at $z = 0$ are slightly less compared with the results known from the other experiments. This can be accounted as an indication that probably we need to have a non-gravitational interaction in order to explain BOSS experiment result for $z=2.34$. On the other hand, if new missions will give lower value for the Hubble parameter for $z=0$ consistent with the results obtained here~(but the result for $z=2.34$ will still valid), then the models considered here can be accounted as viable models. In summary, the study of the models of this subsection showed that they can explain the accelerated expansion and can easily give results in an agreement with BOSS experiment. However, the values of $H_{0}$ for the models are not consistent with reported results existing in the literature~(Table~\ref{tab:Table1}). The subject of next two subsections is to demonstrate that non-linear logarithmic non-gravitational interactions are can fix the problem and give results in well agreement with BOSS and other experiments.

\begin{table}[ht!]
  \centering
    \begin{tabular}{ | l | l | l | l | p{2cm} |}
    \hline
 $(H_{0}, A , k, n)$ & $Omh^{2}(z_{1},z_{2})$ & $Omh^{2}(z_{1},z_{3})$ & $Omh^{2}(z_{2},z_{3})$  & $H(z=2.34)$\\
      \hline
 $(0.635, -0.9, 0.15, -2.0) $ & $0.115$ & $0.121$ & $0.122$ & $2.19$ \\
          \hline
 $(0.63, -0.9, 0.0, -2.0)$ & $0.122$ & $0.121$ & $0.121$  & $2.18$\\

          \hline
             
    \end{tabular}
\caption{Estimated values of the parameters of the models and appropriate value for the Hubble parameter at $z=2.34$ with non - interacting varying polytropic and polytropic dark fluids, respectively. To obtain the presented constraints $Omh^{2}$ analysis had been used.}
  \label{tab:Table1}
\end{table}

\subsection{First group of interacting models}

\begin{figure}[h!]
 \begin{center}$
 \begin{array}{cccc}
\includegraphics[width=80 mm]{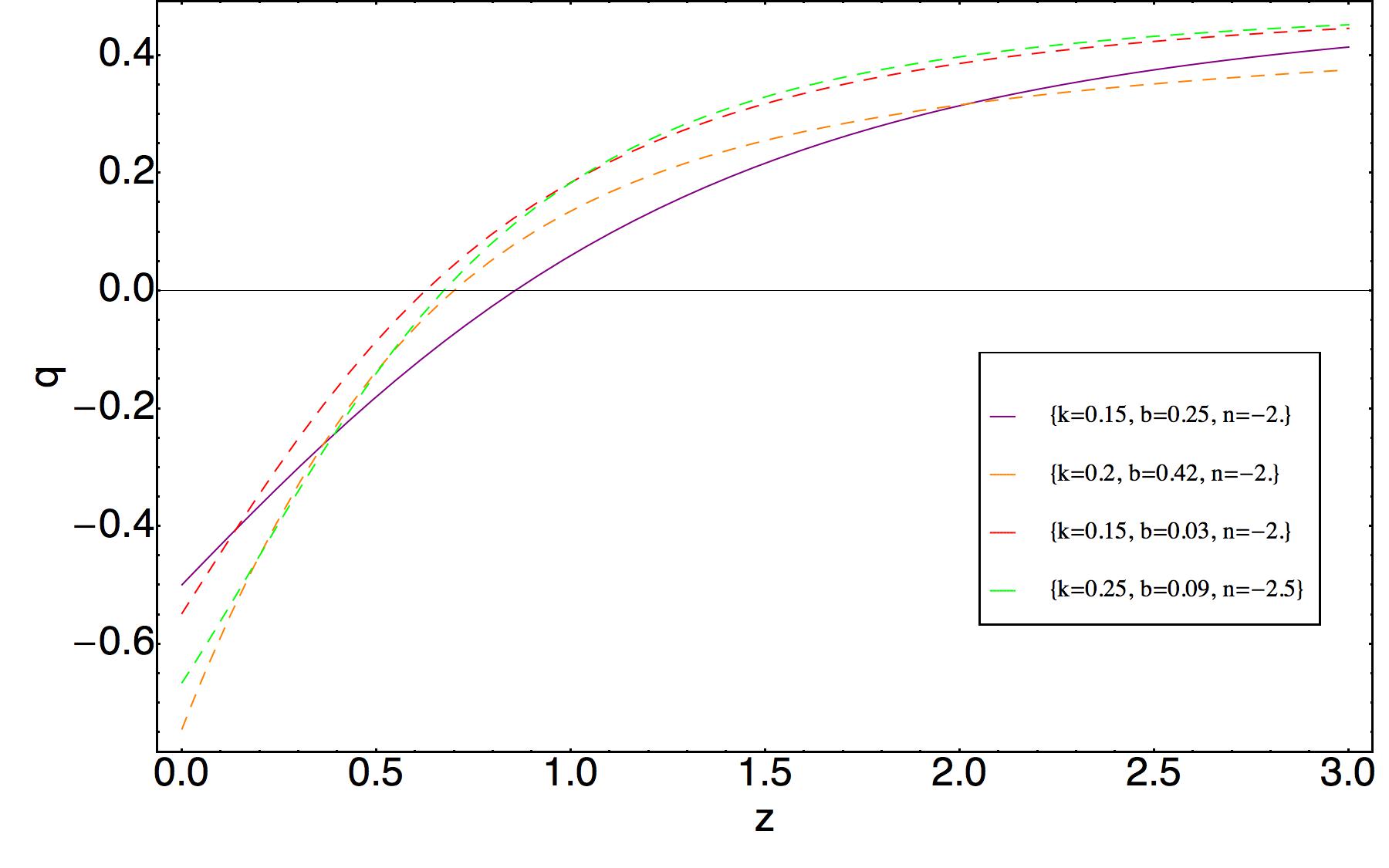} &
\includegraphics[width=80 mm]{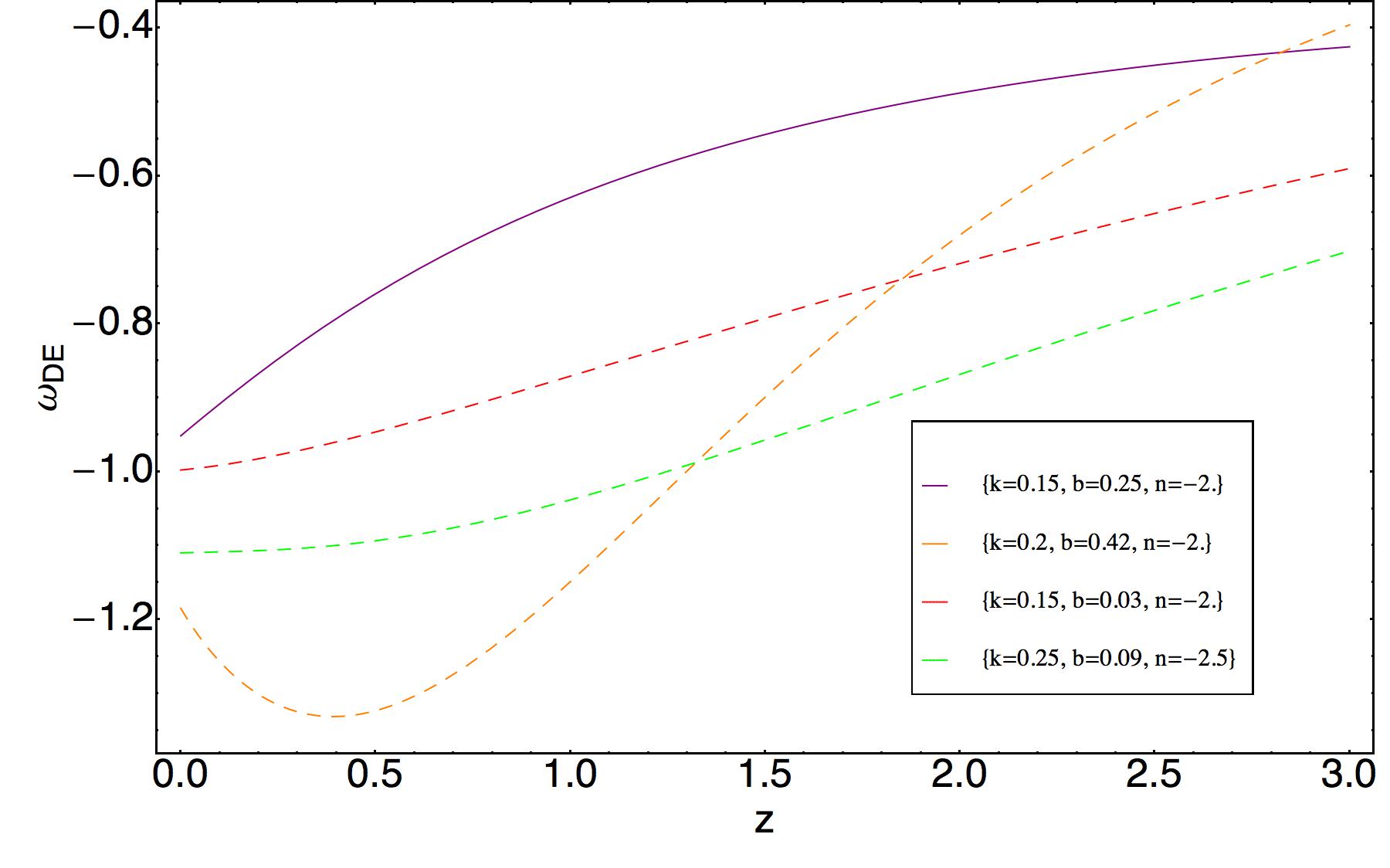}  \\
\includegraphics[width=80 mm]{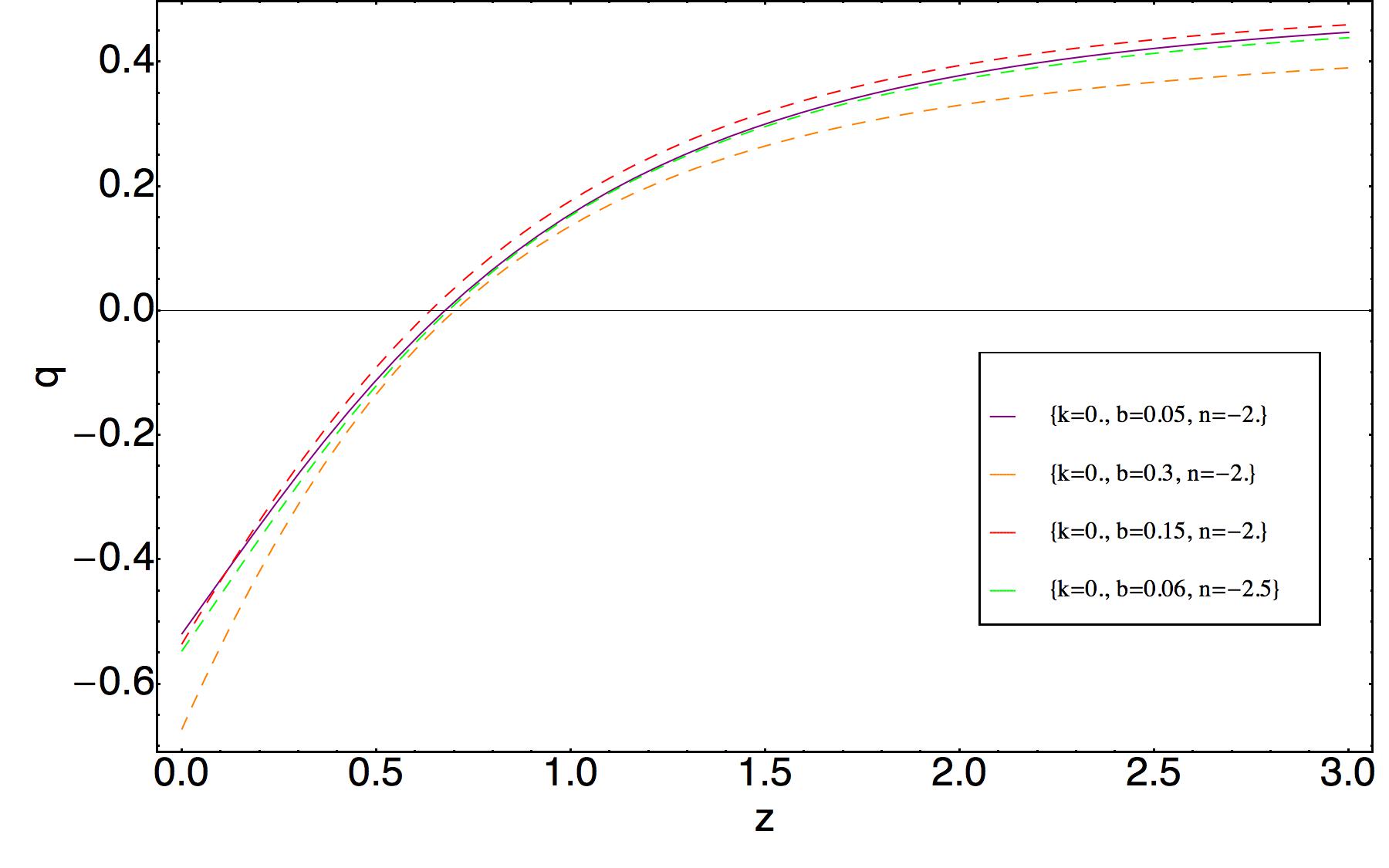} &
\includegraphics[width=80 mm]{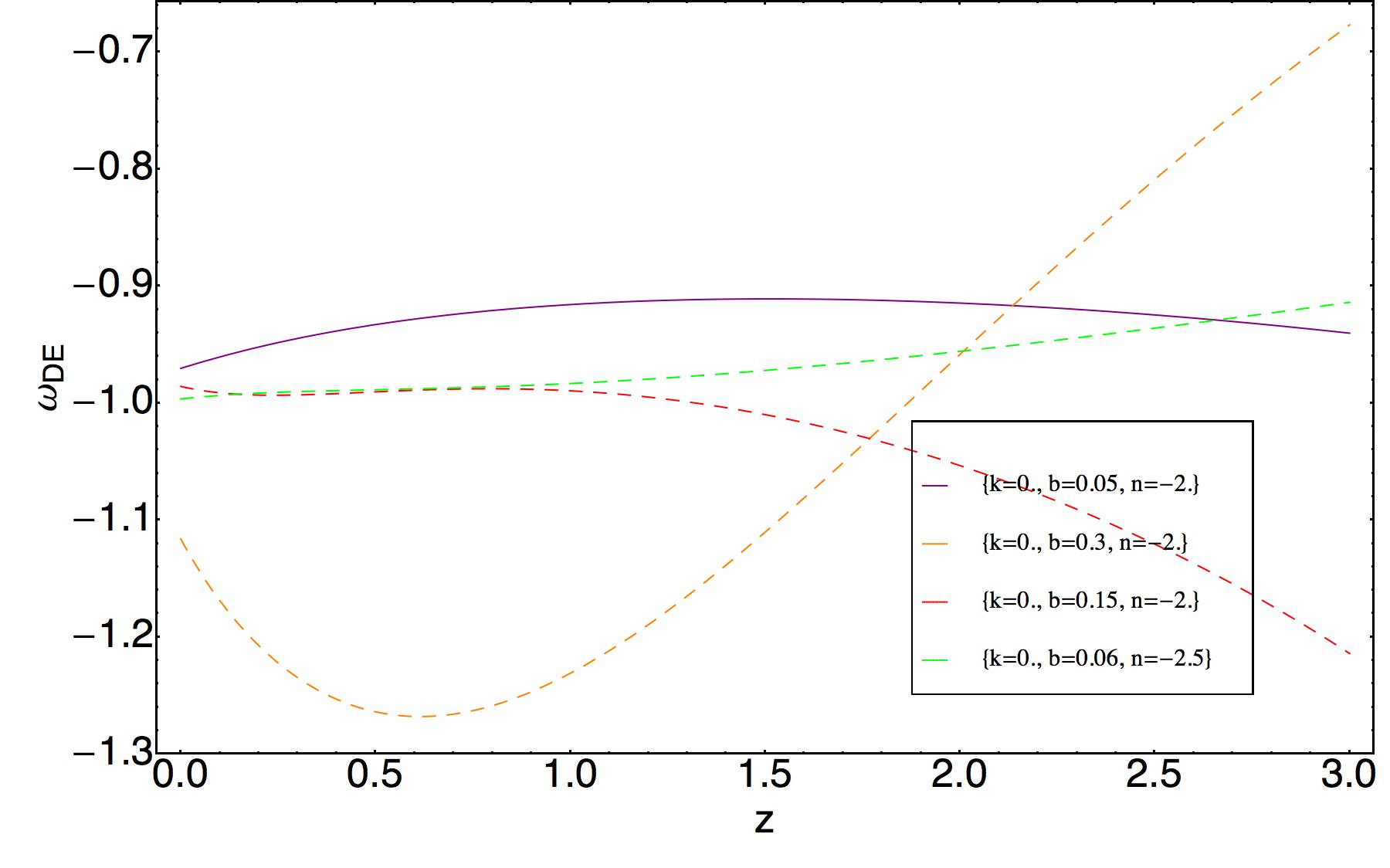}  \\
 \end{array}$
 \end{center}
\caption{The graphical behavior of the deceleration parameter $q$ and the equation of state parameter $\omega_{de}$ against the redshift $z$. The top panel represents the graphical behavior of mentioned parameters for the cosmological models with non-gravitationally interacting varying polytropic dark fluid models. The bottom panel of the plots represents the graphical behavior of the same parameters for the cosmological models with non-gravitationally interacting polytropic dark fluid models. In both cases the interactions are given by Eq.~(\ref{eq:Q1})~(purple curve), Eq.~(\ref{eq:Q2})~(orange curve), Eq.~(\ref{eq:Q5})~(red curve) and Eq.~(\ref{eq:Q6})~(green curve), respectively.}
 \label{fig:Fig3}
\end{figure}

The purpose of this subsection is to demonstrate that non-gravitational interactions of a specific form can be supported by various observational data, can explain the accelerated expansion and provide a solution to the cosmological coincidence problem. In the first group we included the models, where the non-gravitational interactions have been obtained from the following general case
\begin{equation}
Q = 3 H b q^{m} \rho_{de} \log \left [ \frac{\rho_{i}}{\rho_{j}} \right ],
\end{equation}
where $H$ is the Hubble parameter, $b$ is the constant and defines the speed of the energy transition between dark energy and dark matter depends on the sign of it, while $i$ and $j$ can take two values indicating either the energy density of dark energy, or dark matter. On the other hand, parameter $m$ has been introduce allowing to obtain sign changeable interactions for $b>0$ via the deceleration parameter $q$. In general, in this subsection we will consider following $4$ forms of non-gravitational interactions

\begin{equation}\label{eq:Q1}
Q = 3 H b \rho_{de} \log \left[ \frac{\rho_{de}}{\rho_{dm}}\right],
\end{equation}

\begin{equation}\label{eq:Q2}
Q = 3 H b \rho_{de} \log \left[ \frac{\rho_{dm}}{\rho_{de}}\right],
\end{equation}

\begin{equation}\label{eq:Q5}
Q = 3 H b q \rho_{de} \log \left[ \frac{\rho_{de}}{\rho_{dm}}\right],
\end{equation}

\begin{equation}\label{eq:Q6}
Q = 3 H b q \rho_{de} \log \left[ \frac{\rho_{dm}}{\rho_{de}}\right]
\end{equation}
and discuss appropriate consequences relevant to the cosmology of the large scale Universe. The graphical behavior of the deceleration parameter $q$ presented in Fig.~(\ref{fig:Fig3}) clearly indicate that considered forms of non-gravitational interactions Eq.~(\ref{eq:Q1}), Eq.~(\ref{eq:Q2}), Eq.~(\ref{eq:Q5}) and Eq.~(\ref{eq:Q6}) between varying polytropic fluid, Eq.~(\ref{eq:VP}), can be used in order to explain the accelerated expansion of the large scale Universe. 

\begin{figure}[h!]
 \begin{center}$
 \begin{array}{cccc}
\includegraphics[width=80 mm]{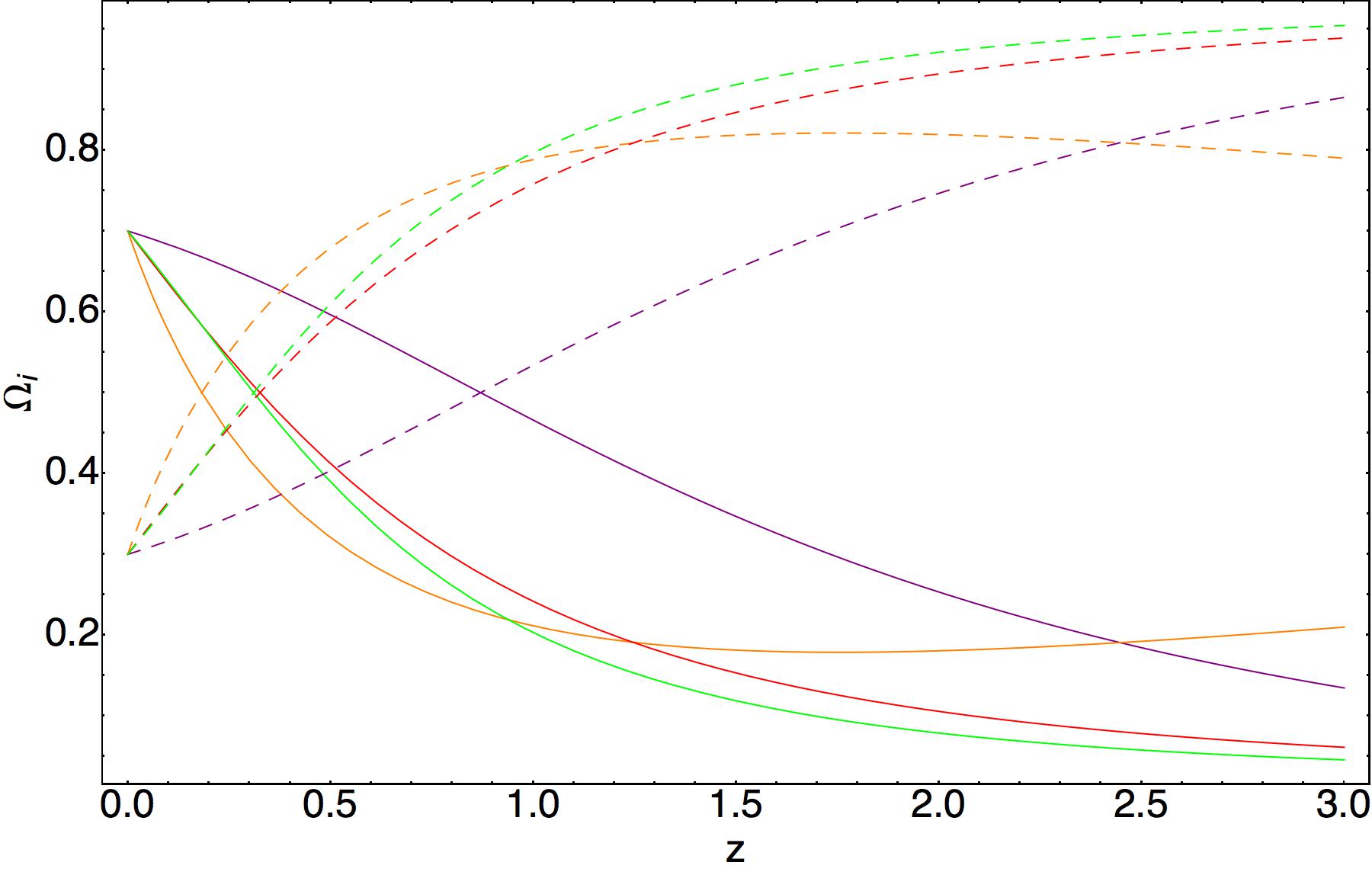} &
\includegraphics[width=80 mm]{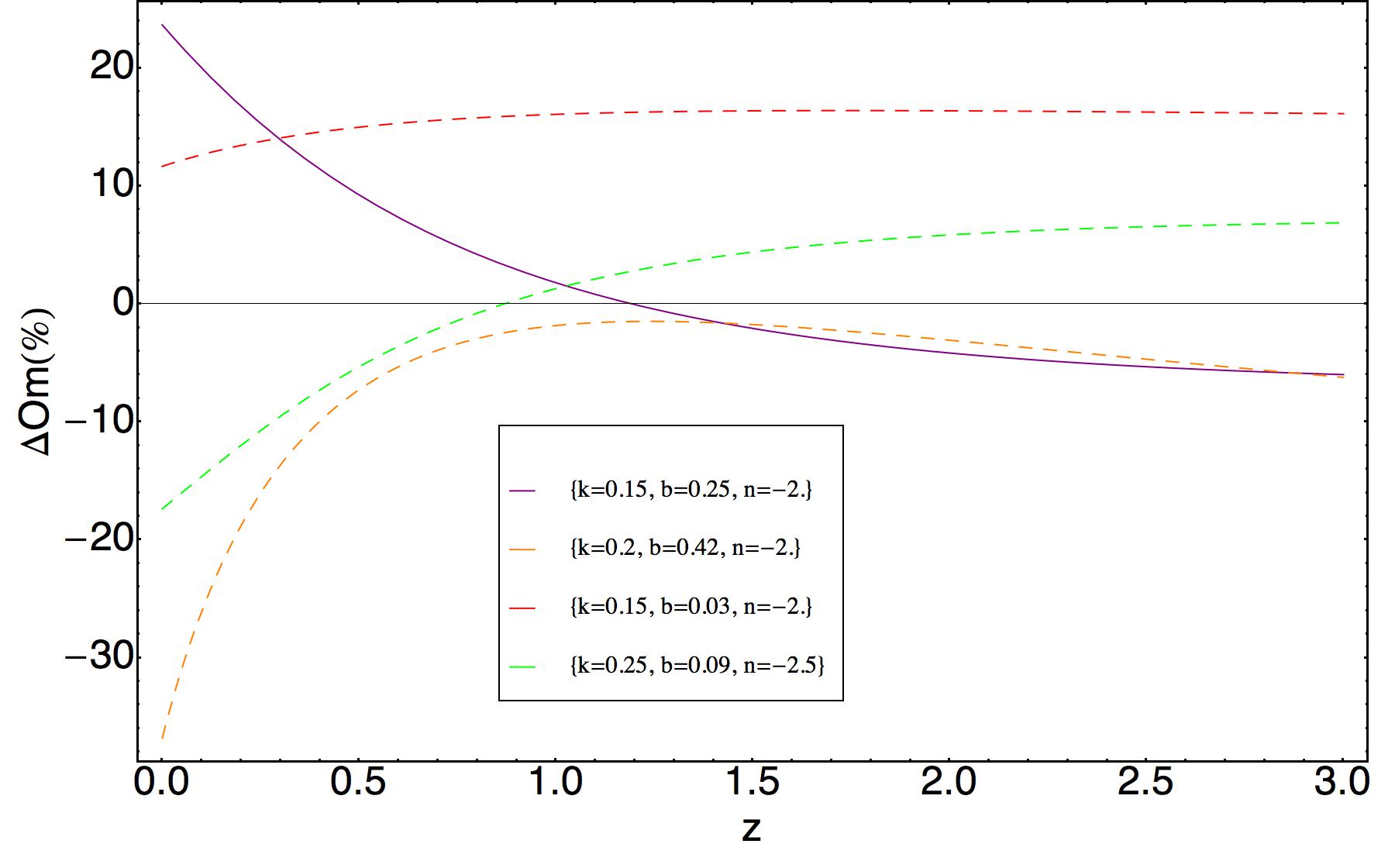} \\
\includegraphics[width=80 mm]{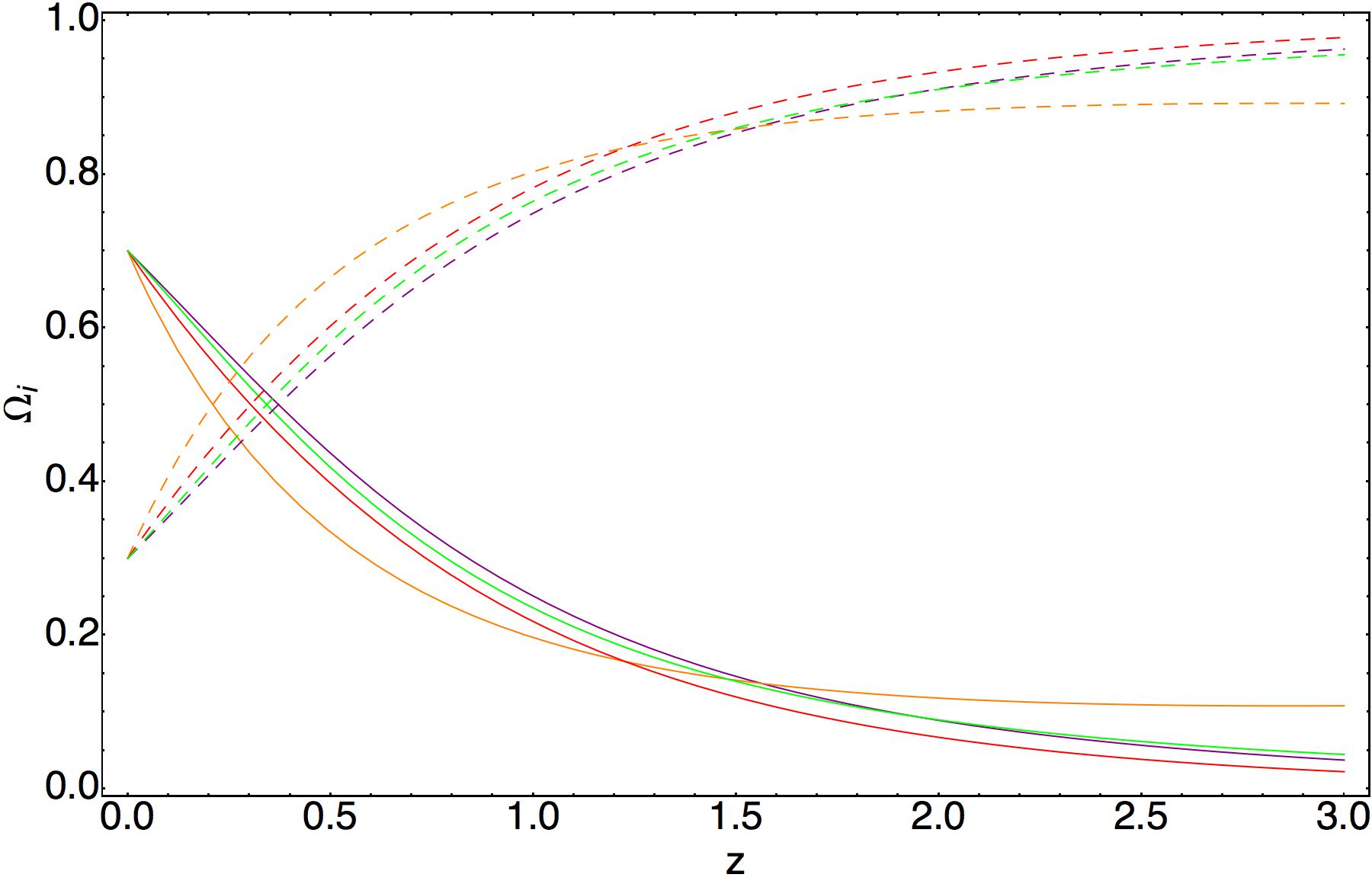} &
\includegraphics[width=80 mm]{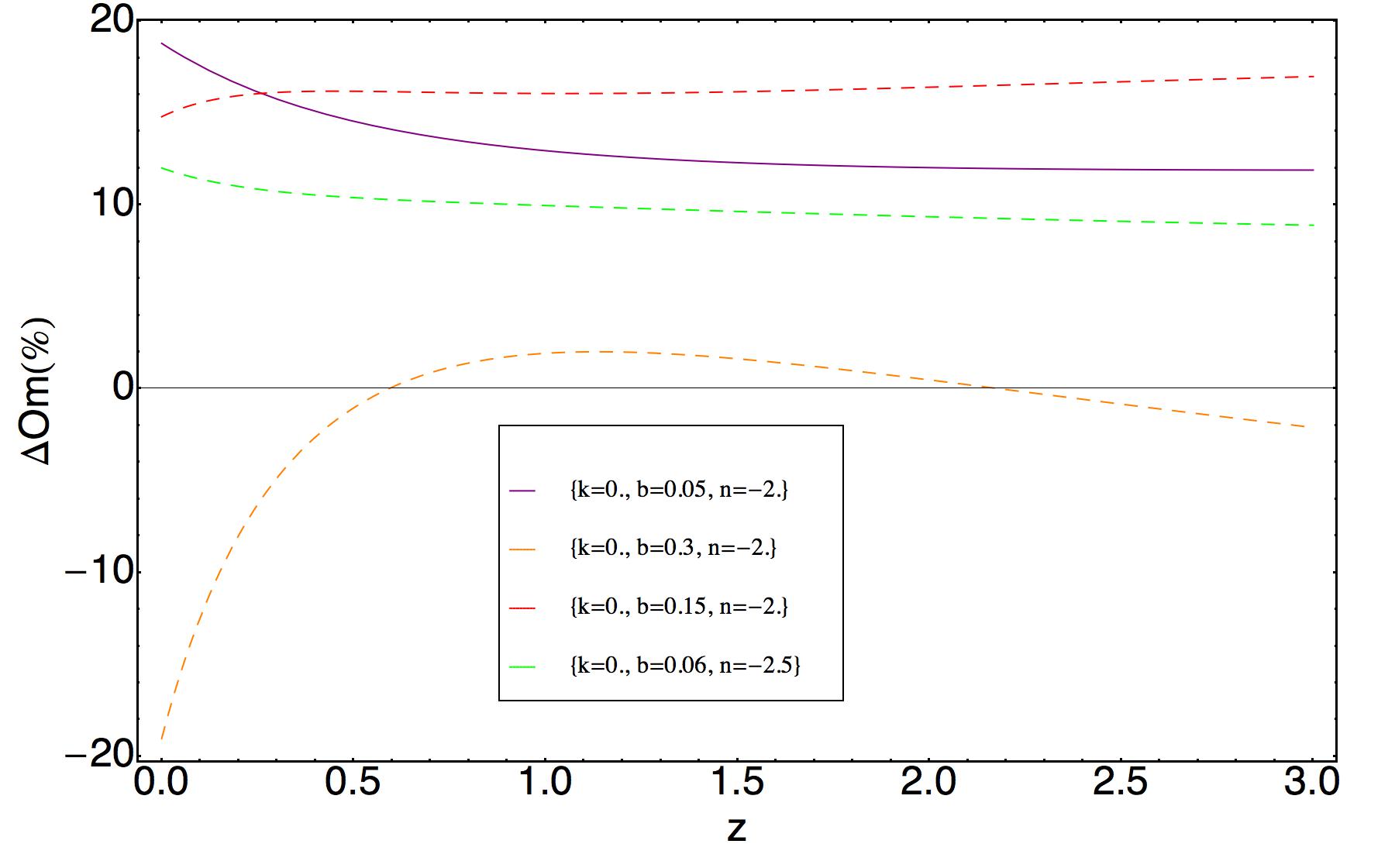} \\  
 \end{array}$
 \end{center}
\caption{The graphical behavior of $\Omega_{de}$ and $\Omega_{dm}$ parameters with $\Delta Om$ against the redshift $z$. The top panel represents the graphical behavior of mentioned parameters for the cosmological models with non-gravitationally interacting varying polytropic dark fluid models. The bottom panel of the plots represents the graphical behavior of the same parameters for the cosmological models with non-gravitationally interacting polytropic dark fluid models. Presented behavior for $\Omega_{de}$ and $\Omega_{dm}$ is according to the same values of the parameters as for the behavior of the deceleration parameter $q$. The solid lines on $\Omega_{i} - z$ plane represent the behavior of $\Omega_{de}$, while dashed lines represent the behavior of $\Omega_{dm}$. In both cases the interactions are given by Eq.~(\ref{eq:Q1})~(purple curve), Eq.~(\ref{eq:Q2})~(orange curve), Eq.~(\ref{eq:Q5})~(red curve) and Eq.~(\ref{eq:Q6})~(green curve), respectively.
}
 \label{fig:Fig4}
\end{figure}

Moreover, the same plot shows, that different models have different transitions redshifts, which are in well correspondence with other estimations existing in literature. On the other hand, the graphical behavior of the EoS parameter presented in the right plot of the top panel of Fig.~(\ref{fig:Fig3}), shows that a combination of presented theoretical results with the results available from PLANCK 2015 experiment, only non-gravitational interactions given by Eq.~(\ref{eq:Q1})~(purple curve) and Eq.~(\ref{eq:Q5})~(red curve) will be supported. This is due to the constraints imposed on the EoS parameter. On the other hand, the graphical behavior of the deceleration and EoS parameters presented on the bottom panel of Fig.~(\ref{fig:Fig3}) corresponding to non-gravitationally interacting polytropic dark fluid, explores several interesting aspects differ compared with the other model. In particular, we see that in addition to imposed constraints if we take into account constraints on the EoS parameter from PLANCK 2015 experiment, then the interactions given by Eq.~(\ref{eq:Q1})~(purple curve), Eq.~(\ref{eq:Q5})~(red curve) and Eq.~(\ref{eq:Q6})~(green curve) will be supported. An interesting solution to the accelerated expansion problem is provided by the interaction Eq.~(\ref{eq:Q5})~(red curve) - it provides a unification of early phantom phase with recent quintessence expanding phase. The graphical behaviors of $\Omega_{de}$ and $\Omega_{dm}$ parameters with the relative change of $Om$ parameter defined by Eq.~(\ref{eq:DeltaOm}) are presented in Fig.~(\ref{fig:Fig4}) for both models with interacting varying and usual polytropic dark fluids.

In Table~\ref{tab:Table2} and Table~\ref{tab:Table3} we collected the best fit values of the parameters obtained from $Omh^{2}$ analysis for interacting varying polytropic and usual polytropic dark fluids, respectively. The last column of these tables represents estimated values of the Hubble parameter for $z=2.34$ according to suggested new models. In summary, in this subsection we demonstrated that BOSS experiment result for the Hubble parameter can be explained as well, when appropriate forms of non-linear logarithmic interactions are exist between dark energy like fluid and cold dark matter. On the other hand, if we will take into account the reported values of $H_{0}$, then we see, that the only viable model is the model with varying polytropic fluid and non-gravitational interaction Eq.~(\ref{eq:Q1}). The status of the other models depends on new observational data.    

\begin{table}[ht!]
  \centering
    \begin{tabular}{ | l | l | l | l | l | p{2cm} |}
    \hline
$Q$ & $(H_{0}, A , k, n, b)$ & $Omh^{2}(z_{1},z_{2})$ & $Omh^{2}(z_{1},z_{3})$ & $Omh^{2}(z_{2},z_{3})$  & $H(z=2.34)$\\
      \hline
Eq.~(\ref{eq:Q1}) &  $(0.69, -0.90, 0.15, -2.0, 0.25) $ & $0.139$ & $0.122$ & $0.121$ & $2.22$ \\
          \hline
Eq.~(\ref{eq:Q2}) &  $(0.69, -1.10, 0.20, -2.0, 0.42)$ & $0.121$ & $0.123$ & $0.123$  & $2.22$\\

          \hline
          
Eq.~(\ref{eq:Q5}) &  $(0.63, -0.85, 0.15, -2.0, 0,03)$ & $0.123$ & $0.125$ & $0.125$  & $2.22$\\

          \hline
          
Eq.~(\ref{eq:Q6}) &  $(0.65, -0.95, 0.25, -2.5, 0.09)$ & $0.110$ & $0.121$ & $0.122$  & $2.20$\\

          \hline                    
             
    \end{tabular}
\caption{Estimated values of the parameters of the models and appropriate value for the Hubble parameter at $z=2.34$ with non - gravitationally interacting varying polytropic dark fluids. To obtain the presented constraints $Omh^{2}$ analysis had been used. The interactions are given by Eq.~(\ref{eq:Q1}), Eq.~(\ref{eq:Q2}), Eq.~(\ref{eq:Q5}) and Eq.~(\ref{eq:Q6}).}
  \label{tab:Table2}
\end{table}
      
\begin{table}[ht!]
  \centering
    \begin{tabular}{ | l | l | l | l | l | p{2cm} |}
    \hline
$Q$ & $(H_{0}, A , k, n, b)$ & $Omh^{2}(z_{1},z_{2})$ & $Omh^{2}(z_{1},z_{3})$ & $Omh^{2}(z_{2},z_{3})$  & $H(z=2.34)$\\
      \hline
Eq.~(\ref{eq:Q1}) &  $(0.64, -0.90, 0.0, -2.0, 0.05) $ & $0.126$ & $0.124$ & $0.123$ & $2.21$ \\
          \hline
Eq.~(\ref{eq:Q2}) &  $(0.68, -1.10, 0.0, -2.0, 0.30)$ & $0.124$ & $0.124$ & $0.124$  & $2.23$\\

          \hline
          
Eq.~(\ref{eq:Q5}) &  $(0.63, -0.90, 0.0, -2.0, 0.15)$ & $0.124$ & $0.125$ & $0.125$  & $2.22$\\

          \hline
          
Eq.~(\ref{eq:Q6}) &  $(0.65, -0.95, 0.0, -2.5, 0.06)$ & $0.126$ & $0.125$ & $0.124$  & $2.22$\\

          \hline                    
             
    \end{tabular}
\caption{Estimated values of the parameters of the models and appropriate value for the Hubble parameter at $z=2.34$ with non - gravitationally interacting polytropic dark fluids. To obtain the presented constraints $Omh^{2}$ analysis had been used. The interactions are given by Eq.~(\ref{eq:Q1}), Eq.~(\ref{eq:Q2}), Eq.~(\ref{eq:Q5}) and Eq.~(\ref{eq:Q6}).}
  \label{tab:Table3}
\end{table}

\subsection{Second group of interacting models}

\begin{figure}[h!]
 \begin{center}$
 \begin{array}{cccc}
\includegraphics[width=80 mm]{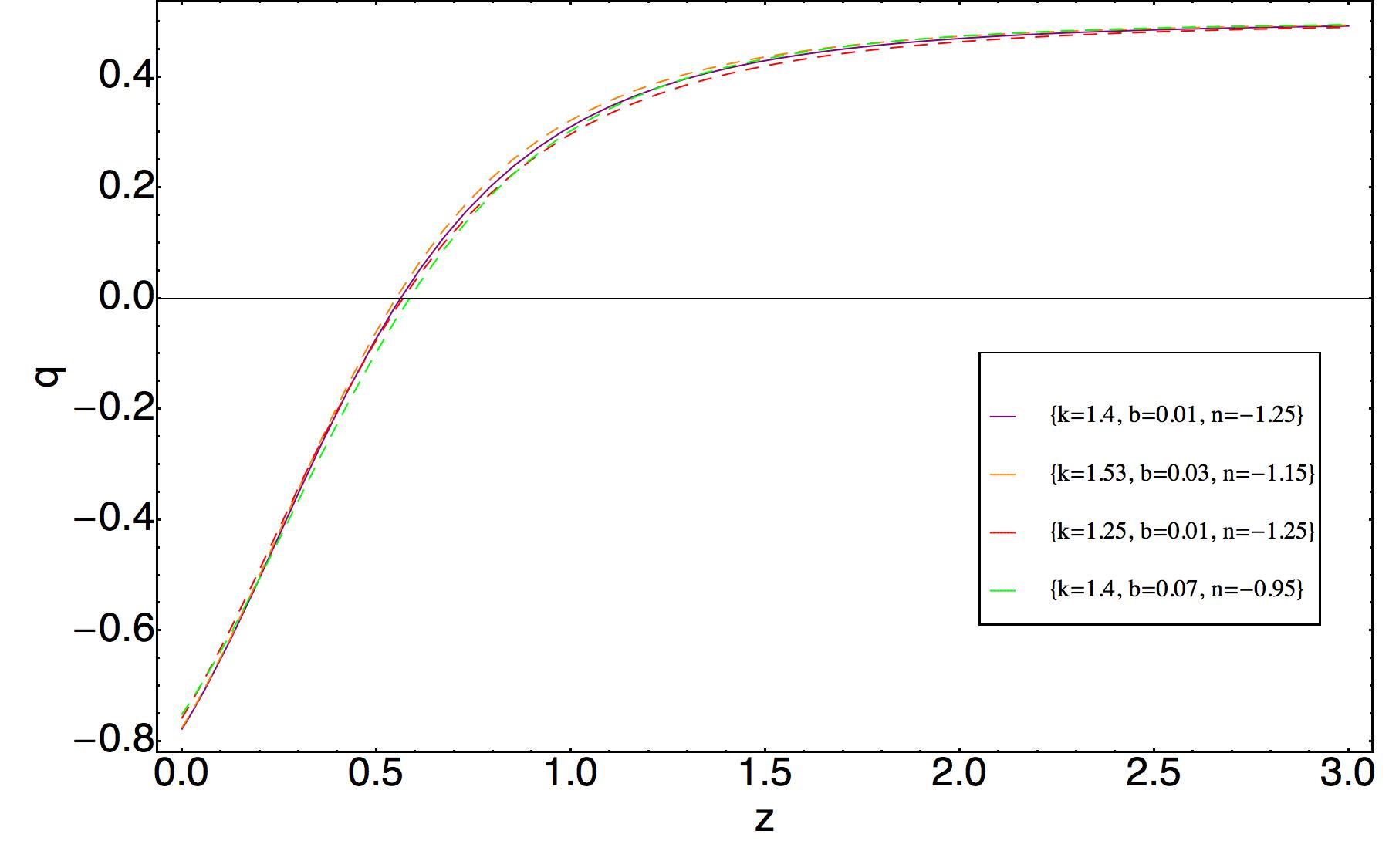} &
\includegraphics[width=80 mm]{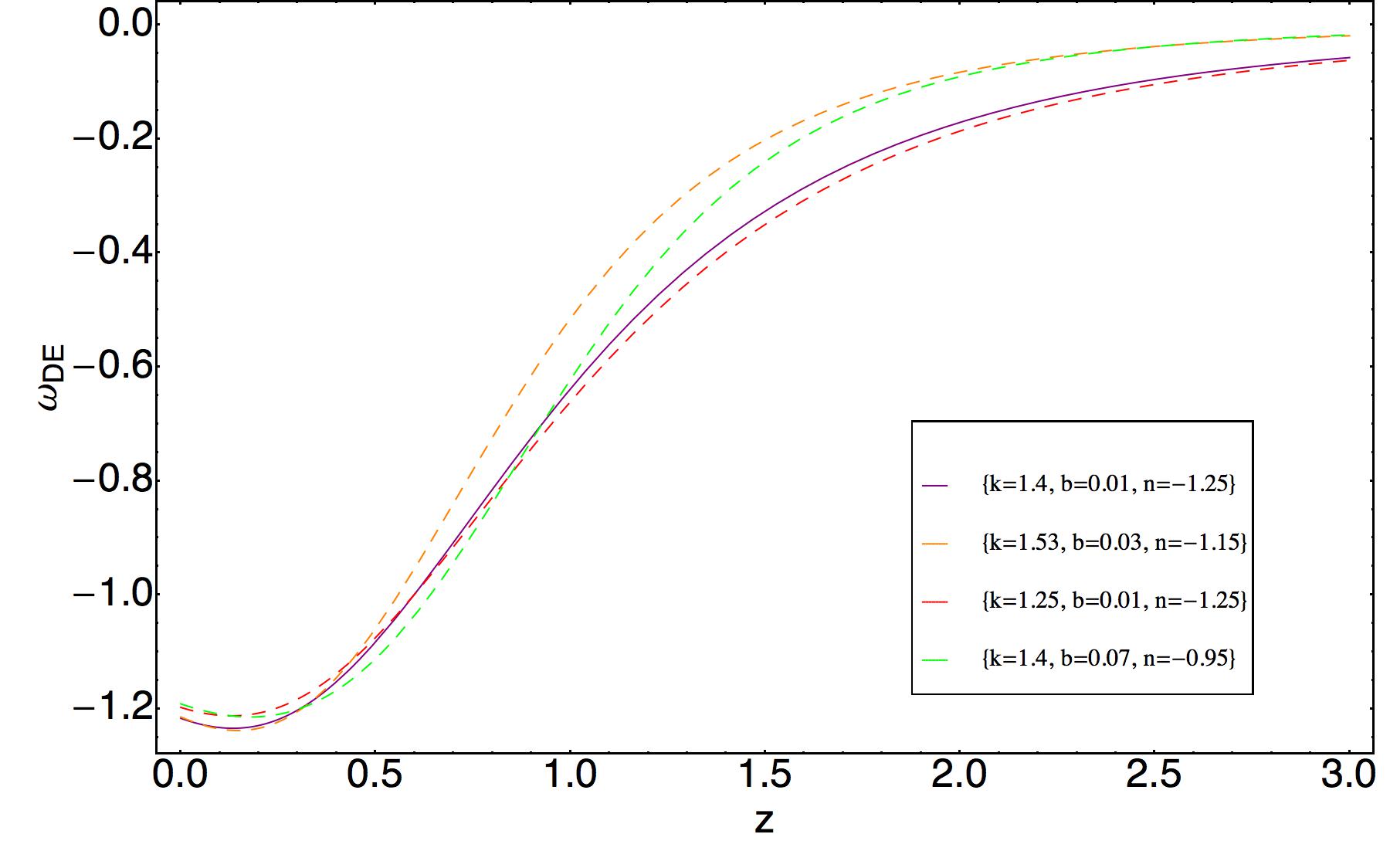}  \\
\includegraphics[width=80 mm]{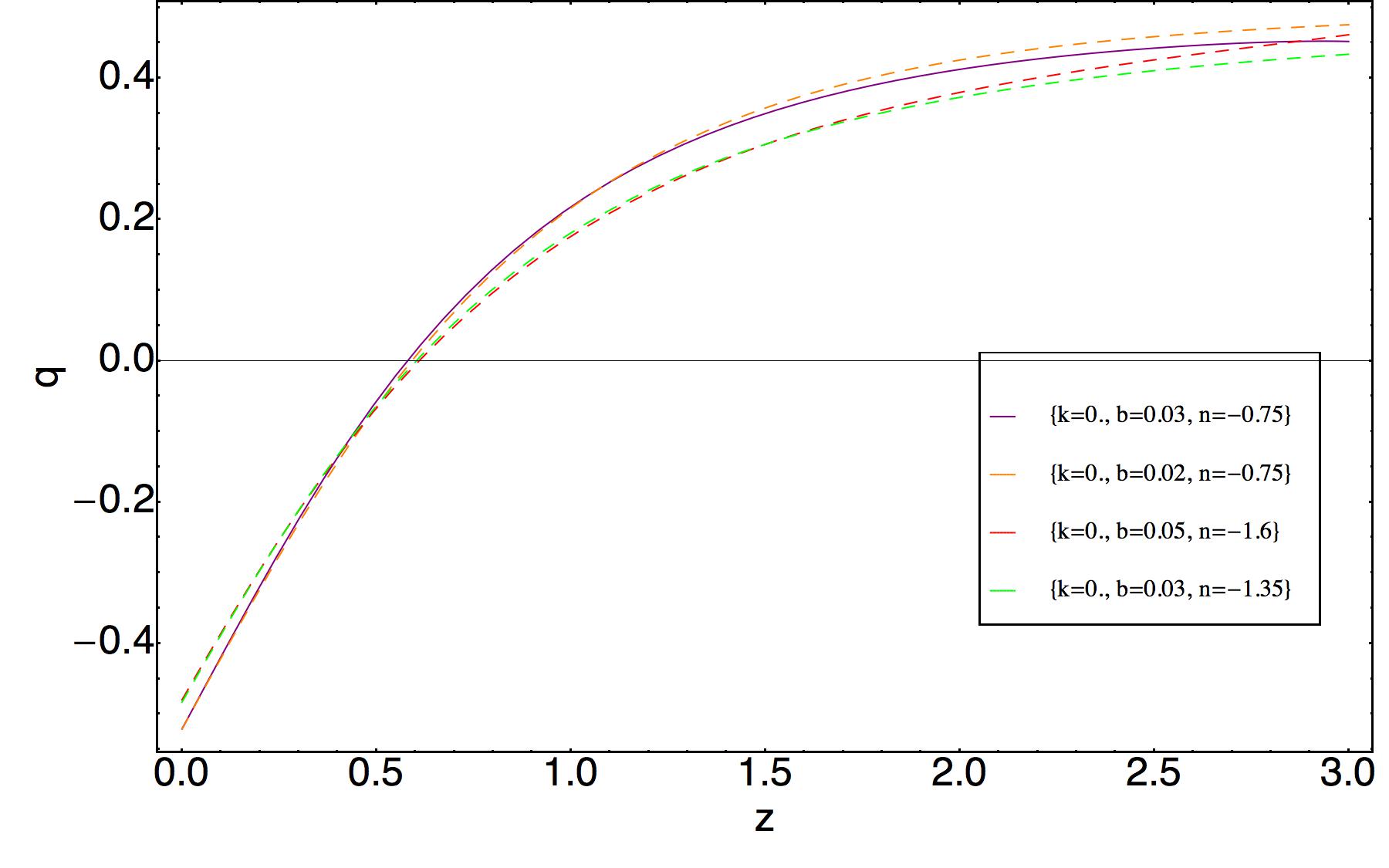} &
\includegraphics[width=80 mm]{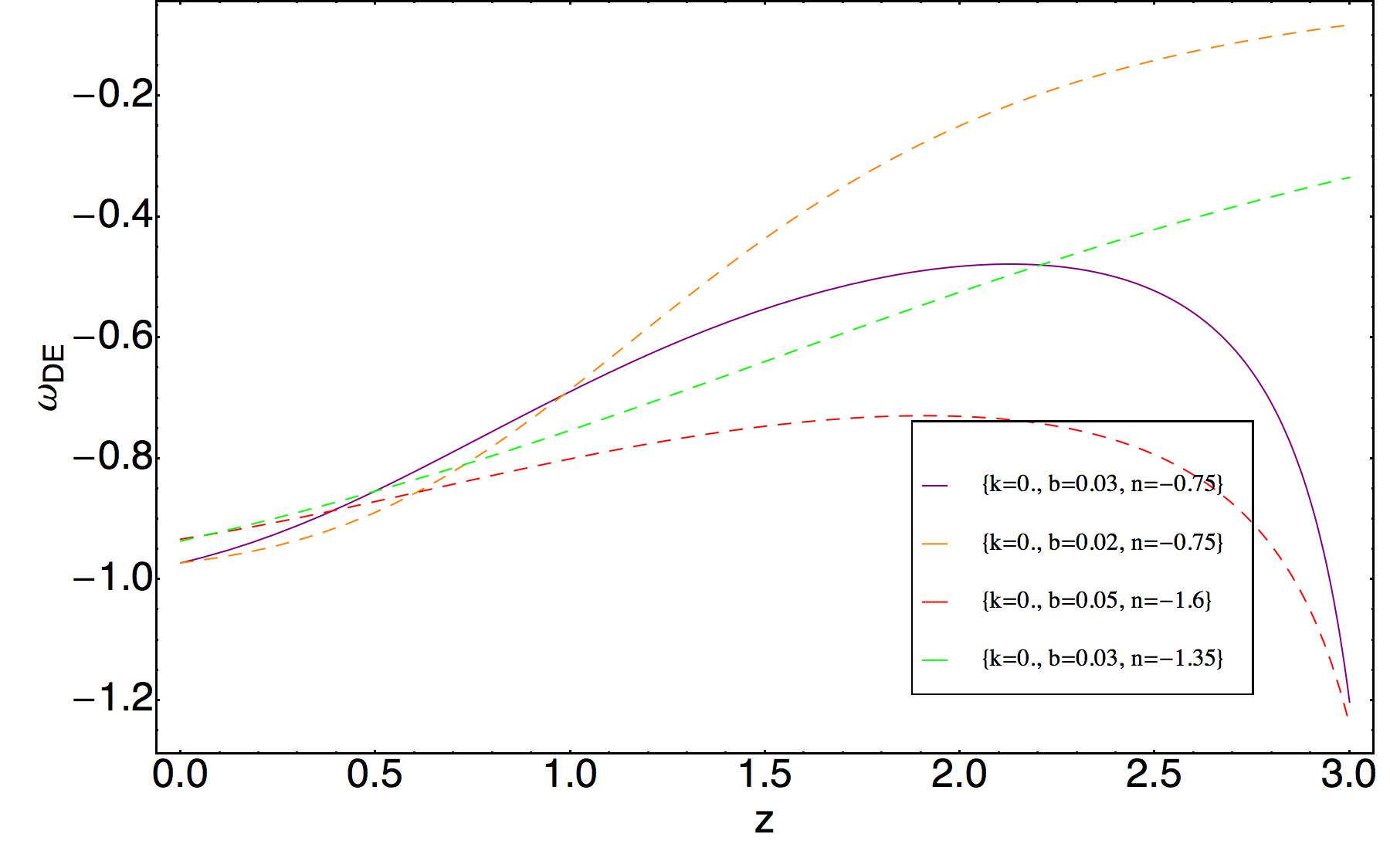}  \\
 \end{array}$
 \end{center}
\caption{The graphical behavior of the deceleration parameter $q$ and the equation of state parameter $\omega_{de}$ against the redshift $z$. The top panel represents the graphical behavior of mentioned parameters for the cosmological models with non-gravitationally interacting varying polytropic dark fluid models. The bottom panel of the plots represents the graphical behavior of the same parameters for the cosmological models with non-gravitationally interacting polytropic dark fluid models. In both cases the interactions are given by Eq.~(\ref{eq:Q3})~(purple curve), Eq.~(\ref{eq:Q4})~(orange curve), Eq.~(\ref{eq:Q7})~(red curve) and Eq.~(\ref{eq:Q8})~(green curve), respectively.}
 \label{fig:Fig5}
\end{figure}

\begin{figure}[h!]
 \begin{center}$
 \begin{array}{cccc}
\includegraphics[width=80 mm]{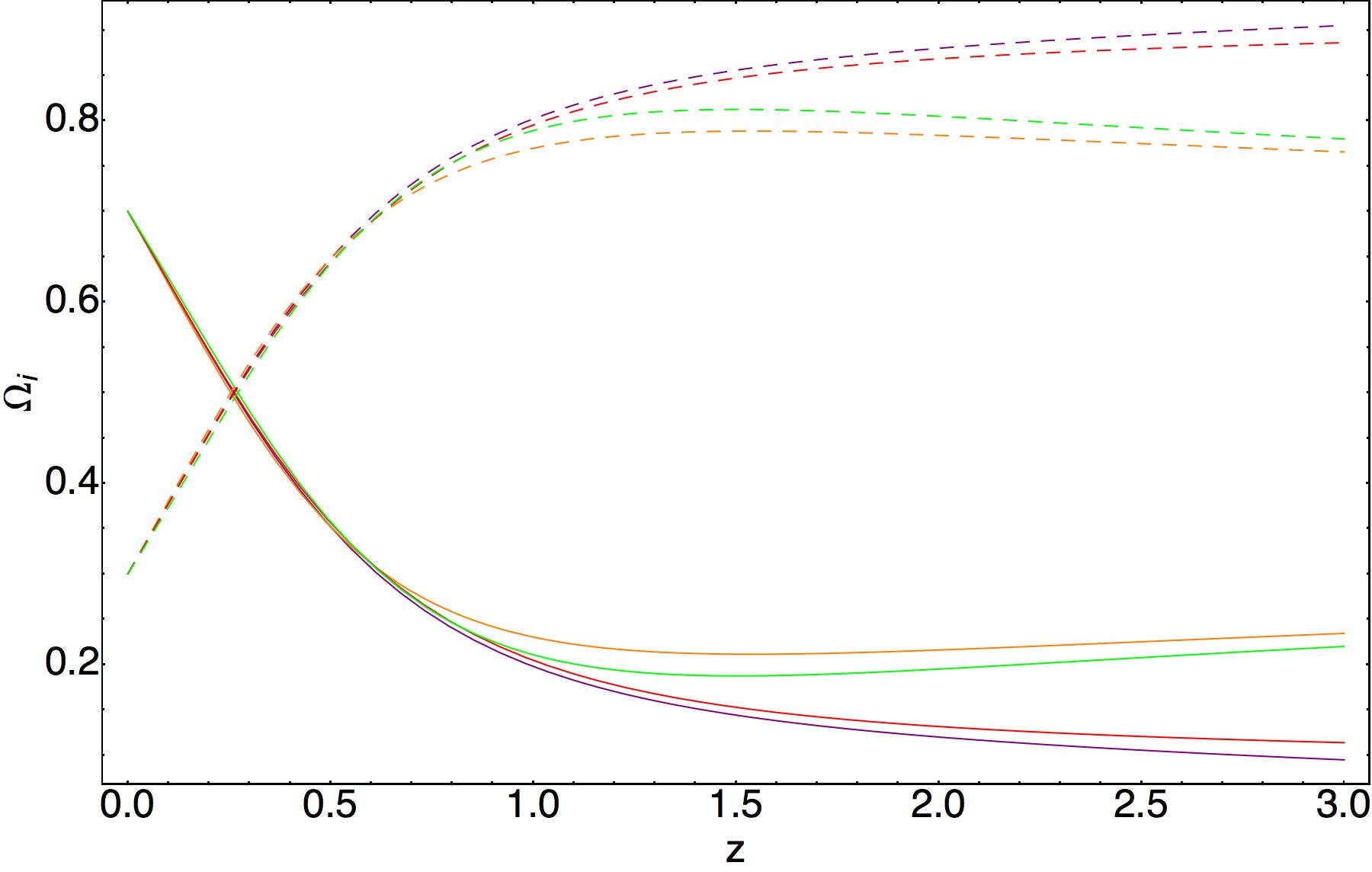} &
\includegraphics[width=80 mm]{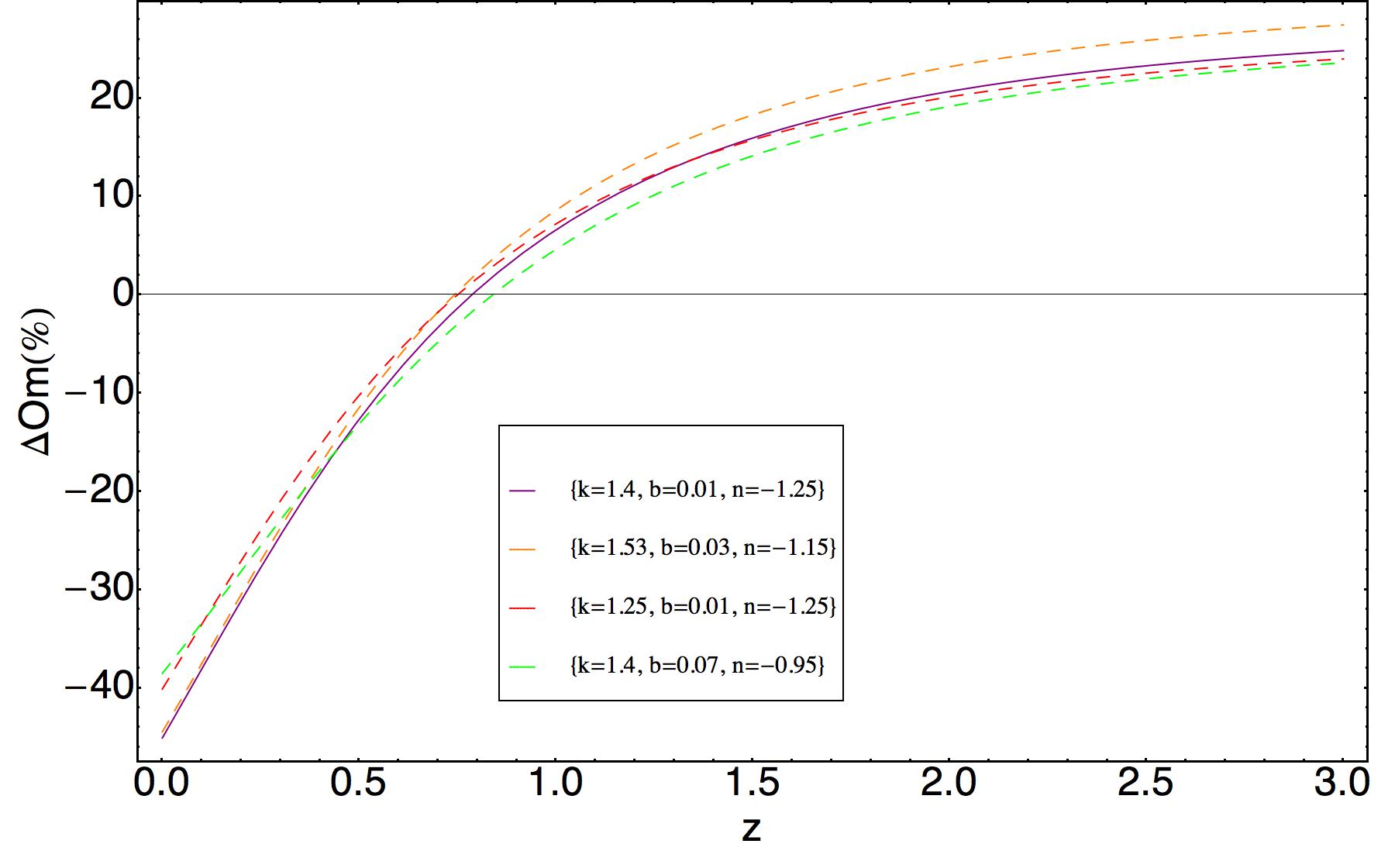} \\
\includegraphics[width=80 mm]{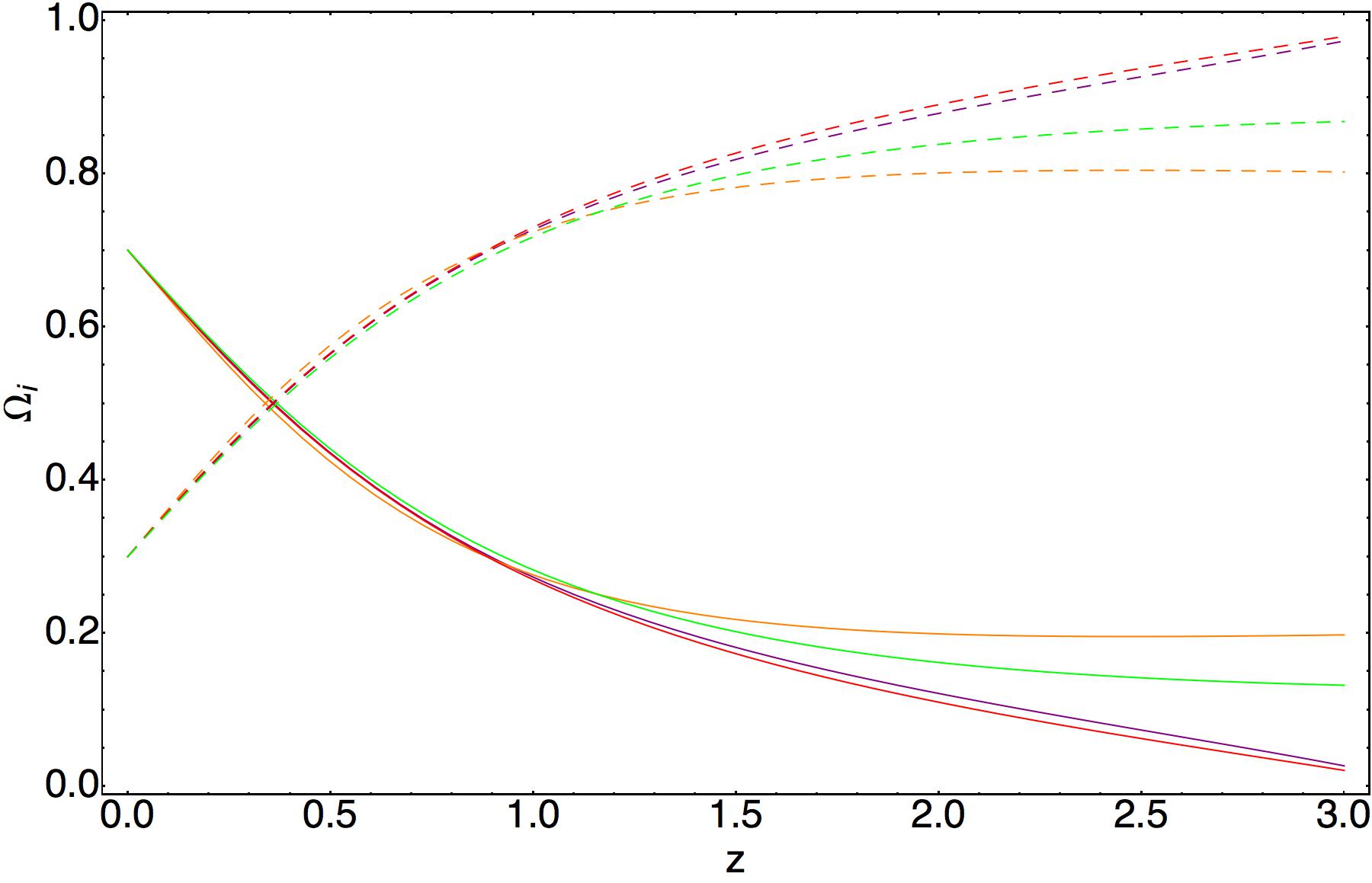} &
\includegraphics[width=80 mm]{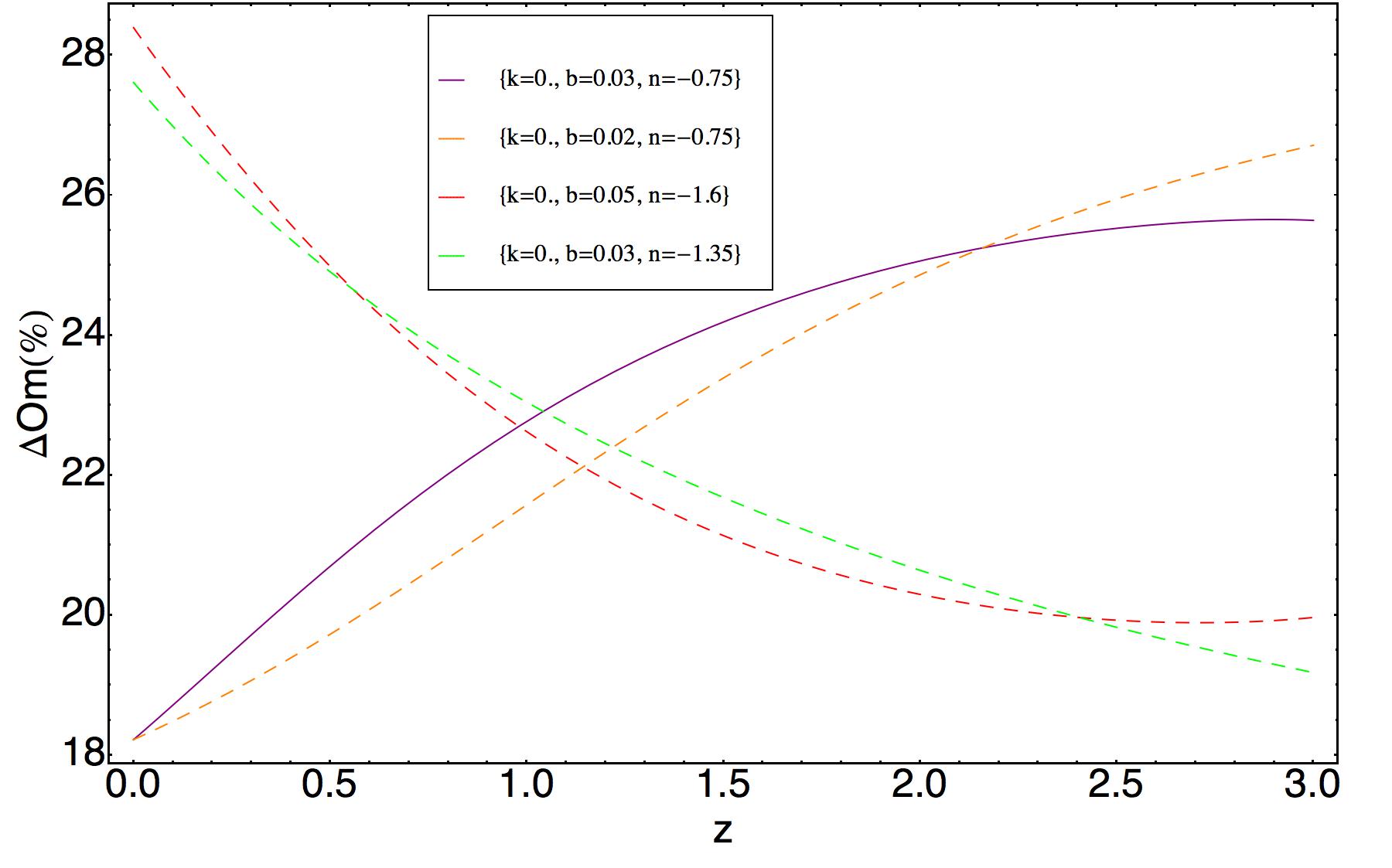} \\  
 \end{array}$
 \end{center}
\caption{The graphical behavior of $\Omega_{de}$ and $\Omega_{dm}$ parameters with $\Delta Om$ against the redshift $z$. The top panel represents the graphical behavior of mentioned parameters for the cosmological models with non-gravitationally interacting varying polytropic dark fluid models. The bottom panel of the plots represents the graphical behavior of the same parameters for the cosmological models with non-gravitationally interacting polytropic dark fluid models. Presented behavior for $\Omega_{de}$ and $\Omega_{dm}$ is according to the same values of the parameters as for the behavior of the deceleration parameter $q$. The solid lines on $\Omega_{i} - z$ plane represent the behavior of $\Omega_{de}$, while dashed lines represent the behavior of $\Omega_{dm}$. In both cases the interactions are given by Eq.~(\ref{eq:Q3})~(purple curve), Eq.~(\ref{eq:Q4})~(orange curve), Eq.~(\ref{eq:Q7})~(red curve) and Eq.~(\ref{eq:Q8})~(green curve), respectively.}
 \label{fig:Fig6}
\end{figure}

The following forms of non-gravitational interactions 

\begin{equation}\label{eq:Q3}
Q = 3 H b \rho_{dm} \log \left[ \frac{\rho_{de}}{\rho_{dm}}\right],
\end{equation}

\begin{equation}\label{eq:Q4}
Q = 3 H b \rho_{dm} \log \left[ \frac{\rho_{dm}}{\rho_{de}}\right],
\end{equation}

\begin{equation}\label{eq:Q7}
Q = 3 H b q \rho_{dm} \log \left[ \frac{\rho_{de}}{\rho_{dm}}\right],
\end{equation}

\begin{equation}\label{eq:Q8}
Q = 3 H b q \rho_{dm} \log \left[ \frac{\rho_{dm}}{\rho_{de}}\right],
\end{equation}
have been used in order to construct cosmological models identified as the cosmological models of the second group. First of all, to simplify the discussion and illuminate physically relevant behavior, we used the same constraints as had been used for the models considered in previous subsections. Obtained constraints with the estimated value of the Hubbel parameter at $z=2.34$ are presented in Table~\ref{tab:Table4} and Table~\ref{tab:Table5}, for interacting varying and usual polytropic dark fluids respectively. Then, having the values of the parameters of the models, we studied the behavior of the deceleration parameter $q$, EoS parameter $\omega_{de}$, $\Omega_{de}$ and $\Omega_{dm}$ parameters with the relative change of $Om$ parameter. The results are presented in Fig.~(\ref{fig:Fig5}) and Fig.~(\ref{fig:Fig6}).

\begin{table}[ht!]
  \centering
    \begin{tabular}{ | l | l | l | l | l | p{2cm} |}
    \hline
$Q$ & $(H_{0}, A , k, n, b)$ & $Omh^{2}(z_{1},z_{2})$ & $Omh^{2}(z_{1},z_{3})$ & $Omh^{2}(z_{2},z_{3})$  & $H(z=2.34)$\\
      \hline
Eq.~(\ref{eq:Q3}) &  $(0.61, -0.50, 1.40, -1.25, 0.01) $ & $0.09$ & $0.123$ & $0.126$ & $2.20$ \\
          \hline
Eq.~(\ref{eq:Q4}) &  $(0.61, -0.46, 1.53, -1.15, 0.03)$ & $0.09$ & $0.126$ & $0.129$  & $2.22$\\

          \hline
          
Eq.~(\ref{eq:Q7}) &  $(0.61, -0.53, 1.25, -1.25, 0,01)$ & $0.09$ & $0.122$ & $0.125$  & $2.20$\\

          \hline
          
Eq.~(\ref{eq:Q8}) &  $(0.61, -0.46, 1.40, -0.95, 0.07)$ & $0.09$ & $0.122$ & $0.124$  & $2.20$\\

          \hline                    
             
    \end{tabular}
\caption{Estimated values of the parameters of the models and appropriate value for the Hubble parameter at $z=2.34$ with non - gravitationally interacting varying polytropic dark fluids. To obtain the presented constraints $Omh^{2}$ analysis had been used. The interactions are given by Eq.~(\ref{eq:Q3}), Eq.~(\ref{eq:Q4}), Eq.~(\ref{eq:Q7}) and Eq.~(\ref{eq:Q8}).}
  \label{tab:Table4}
\end{table}
      
\begin{table}[ht!]
  \centering
    \begin{tabular}{ | l | l | l | l | l | p{2cm} |}
    \hline
$Q$ & $(H_{0}, A , k, n, b)$ & $Omh^{2}(z_{1},z_{2})$ & $Omh^{2}(z_{1},z_{3})$ & $Omh^{2}(z_{2},z_{3})$  & $H(z=2.34)$\\
      \hline
Eq.~(\ref{eq:Q3}) &  $(0.61, -0.70, 0.0, -0.75, 0.03) $ & $0.122$ & $0.126$ & $0.126$ & $2.22$ \\
          \hline
Eq.~(\ref{eq:Q4}) &  $(0.61, -0.70, 0.0, -0.75, 0.02)$ & $0.121$ & $0.126$ & $0.127$  & $2.22$\\

          \hline
          
Eq.~(\ref{eq:Q7}) &  $(0.61, -0.80, 0.0, -1.6, 0.05)$ & $0.125$ & $0.121$ & $0.120$  & $2.18$\\

          \hline
          
Eq.~(\ref{eq:Q8}) &  $(0.61, -0.78, 0.0, -1.35, 0.03)$ & $0.125$ & $0.121$ & $0.121$  & $2.18$\\

          \hline                    
             
    \end{tabular}
\caption{Estimated values of the parameters of the models and appropriate value for the Hubble parameter at $z=2.34$ with non - gravitationally interacting polytropic dark fluids. To obtain the presented constraints $Omh^{2}$ analysis had been used. The interactions are given by Eq.~(\ref{eq:Q3}), Eq.~(\ref{eq:Q4}), Eq.~(\ref{eq:Q7}) and Eq.~(\ref{eq:Q8}).}
  \label{tab:Table5}
\end{table}

Similar to the other models, if we will take into account the constraints on $\omega_{de}$ from PLANCK 2015, then interacting varying polytropic dark energy models should be out from future discussion, since the present day value of $\omega_{de}$ is smaller than the lowest limit obtained in PLANCK 2015 experiment. On the other hand, we have interesting results for the models with interacting usual polytropic fluid supported also from PLANCK 2015 experiment. The graphical behavior of the deceleration parameter and $\omega_{de}$ is presented on the bottom panel of Fig.~(\ref{fig:Fig5}). The study shows that the accelerated expansion of the large scale Universe can be explained when the interactions are given by Eq.~(\ref{eq:Q4})~(orange curve) and Eq.~(\ref{eq:Q8})~(green curve). In this case the dark fluid has only quintessence nature. On the other hand, when the interactions are given by Eq.~(\ref{eq:Q3})~(purple curve) and Eq.~(\ref{eq:Q7})~(red curve), the usual polytropic dark fluid has phantom nature for high redshifts and becomes quintessence at lower redshifts. The left plot of the bottom panel of Fig.~(\ref{fig:Fig6}) represents the graphical behaviors of $\Omega_{de}$ and $\Omega_{dm}$ parameters, while the right plot represents the relative change of $Om$ parameter. It allows to understand quantitative and qualitative differences between considered models and $\Lambda$CDM standard model. The behavior of $\omega_{de}$ already had demonstrated that the forms of non-gravitational interactions provide physically different possibilities to explain the accelerated expansion of the Universe which has been confirmed by the behavior of $\Delta Om$ parameter. In summary, we can claim, that we have constructed cosmological models with logarithmic non-gravitational interactions between usual polytropic dark fluid and cold dark matter which can explain observational data and the results for the Hubble parameter at $z = 2.34$~(Table~\ref{tab:Table5}). However, the constraints on $H_{0}$ reported in the literature demonstrates that the models of the second group are not viable models. Of course, the status of these models can be changed with new observational data.

\section{Discussion}\label{sec:Dis}

Due to various phenomenological assumptions the interpretation of available observational data can be not unique. It can be related also to a tension existing between different observational datasets. However, without phenomenological assumptions it is hard to solve the problems, which exist in modern cosmology. One of the phenomenological assumptions is related to the possible existence of non-gravitational interaction between dark energy and dark matter. There are various forms/classes of non-gravitational interactions considered in literature and they can be described as follows: linear, non-linear, linear sign-changeable and non-linear sign-changeable. In this paper we developed new models of non-gravitational interactions belonging to non-linear and non-linear sign changeable interactions. One of the goals of the paper was to demonstrate that interacting varying polytropic fluid can be applied to explain the accelerated expansion of the recent Universe and can explain the result obtained for the Hubble parameter at $z=2.34$. In particular, the study of non-interacting model showed, that the parameter $k$ raised due to the modification of polytropic fluid considered in this paper does not play a crucial role on the transition redshift. However, this parameter significantly affects on the behavior of $\omega_{de}$ indicating differences between the models conteining usual and varying polytropic dark fluids. In particular, the study shows that EoS parameter for the non-interacting polytropic dark fluid~($k=0$) is linearly decreasing function of the redshift $z$ and the fluid has only quintessence nature. On the other hand, the equation of state parameter for the non-interacting varying polytropic dark fluid is a  decreasing function and started from $z \approx 0.3$ becomes a constant~($\omega_{de} \approx - 1.05$). Moreover, the present day value of the EoS parameter is within the constraints obtained from PLANCK 2015 experiments and the models are free from the cosmological coincidence problem. We found that in this case the deceleration parameter with $\Omega_{de}$ and $\Omega_{dm}$ parameters are not enough for deep and conclusive study of the models. Completely another picture have been observed when the models have been studied via $Om$ analysis allowing to estimate differences between $\Lambda$CDM standard model and suggested new models. Then the constraints imposed by $Omh^{2}$ allowed to see that the estimated value of the Hubble parameter at $z=2.34$ for both models are in well correspondence with the results from BOSS experiment, while the obtained values of $H_{0}$ are not in well correspondence with the results known from other experiments. Therefore, the models are not viable. 

Then we organized two separate subsections combining the interactions given by Eq.~(\ref{eq:Q1}), Eq.~(\ref{eq:Q2}), Eq.~(\ref{eq:Q5}) and Eq.~(\ref{eq:Q6}) in the first group, while the models containing interactions given by Eq.~(\ref{eq:Q3}), Eq.~(\ref{eq:Q4}), Eq.~(\ref{eq:Q7}) and Eq.~(\ref{eq:Q8}) to represent the second group. For each case the behavior of the deceleration and equation of state parameter have been studied. Moreover, constraints form $Omh^{2}$ analysis have been applied in order to obtain the best fit values of the parameters of the models. In particular, the study of the models of the first group with interacting varying polytropic fluid supports possible existance of the non-gravitational interaction Eq.~(\ref{eq:Q1}). In this model, dark energy has only quintessence nature for $z \in [0,3]$ with $\omega_{de} \approx -0.95$ and $\Delta Om \approx 23.5\%$ at $z=0$. On the other hand, if we consider usual polytropic dark fluid, then constraints from $Omh^{2}$ and known results from the constraints on the Hubble parameter at $z = 0$, will support the model with non-gravitational interaction given by Eq.~(\ref{eq:Q2}). However, constraints on $\omega_{de}$ from PLANCK 2015 experiment will make the model not viable. The study of the models of the second group, shows that obtained values of the Hubble parameter for $z=0$ are well smaller than any value reported in literature for this parameter at $z=0$. This fact can be accounted to conclude that the models of this group are not viable. On the other hand, the similar conclusion can be achieved if we will take into account the constraints on $\omega_{de}$ from PLANCK 2015~(in case of interacting varying polytropic fluid). On the other hand, the models of the second group with usual polytropic fluid can be accounted as viable, if we ignore obtained lower values of $H_{0}$ differ from the reported values in the literature. The study of $\omega_{de}$ for these models shows that they can be devided into two sub-classes with quintessence - quintessence and phantom - quintessence transitions well satisfying the constraints on this parameter imposed from PLANCK 2015 experiment. 

In summary, in this paper new parameterizations of non-gravitational interactions have been suggested and using observational constraints appropriate viable model has been seperated among the others~(differ from each other by the form of the interaction form). In particular it has been demonstrated that the model with varying polytropic fluid with non-gravitational interaction given by Eq.~(\ref{eq:Q1}) is the viable model. 

In future it is necessary to study the models in the light of strong gravitational lensing data including structure formation problems. Moreover, it is necessary to classify future type singularities for the models where we have phantom polytropic fluid. These are the models without constraints imposed on $\omega_{de}$ from PLANCK 2015 experiment. Moreover, in this paper we demonstrated that it is not necessary that the EoS of dark energy to explain mentioned BOSS experiment result should have a very exotic form, however, still we need to have non-gravitational interaction. Therefore, it is necessary to extend the present work by including viscosity and other models of dark energy fluids to see the possibilities either to exclude the needs to have non-gravitational interactions, or understand the types of these interactions in order to achieve to the final goal.

\section*{\large{Acknowledgments}}
The authors appreciate Prof. K. Urbanowski from Institute of Physics, University of Zielona Gora, for valuable comments/suggestions, also for inviting their attention on Ref.-s~\cite{U1}~-~\cite{U6} during the preparation of the paper.


\begin{thebibliography}{1}

\bibitem{MR}
M. Roos, John Wiley $\&$ Sons, Ltd, ISBN: 978-1-118-92332-0 (2015)

\bibitem{MG1}
S. Nojiri and S. D. Odintsov, Int. J. Geom. Methods Mod. Phys. 04, 115 (2007)

\bibitem{MG2}
S. Nojiri and S. D. Odintsov, Phys. Rept. 505:59-144 (2011)

\bibitem{MG2}
S. Nojiri and S. D. Odintsov, Phys. Rev. D 68, 123512 (2003)

\bibitem{MG3}
S. Nojiri and S. D. Odintsov, Phys. Rev. D 74, 086005 (2006)

\bibitem{MG4}
S. Capozziello et al, Phys. Lett. B 639:135-143 (2006)

\bibitem{MG5}
G. Cognola et al, Phys. Rev. D 77, 046009 (2008)

\bibitem{MG6}
Yi-Fu Cai et al, Rept. Prog. Phys. 79, no. 4, 106901 (2016)

\bibitem{MG7}
J. B. Dent et al, JCAP 1101 (2011) 009

\bibitem{MG8}
S. Nesseris et al, Phys. Rev. D 88, 103010 (2013)

\bibitem{MG9}
G. Kofinas, E. N. Saridakis, Phys. Rev. D 90, 084044 (2014)

\bibitem{MG10}
M. Skugoreva, Phys. Rev. D 91, 044023 (2015)

\bibitem{MG11}
V.K. Oikonomou, E. N. Saridakis, Phys. Rev. D 94, 124005 (2016)

\bibitem{M3}
T. Clifton et al, Physics Reports 513, 1-189 (2012)

\bibitem{M4}
S. Capozziello, M. De Laurentis, Physics Reports 509, 167–321 (2011)

\bibitem{In1}
Kazuharu Bamba, Sergei D. Odintsov, Symmetry 2015, 7, 220-240

\bibitem{In2}
S. Nojiri et al, arXiv:1705.11098

\bibitem{MG12}
S. Chattopadhyay et al, Eur. Phys. J. C 74:3080 (2014)

\bibitem{MK1}
M. Khurshudyan, Mod. Phys. Lett. A, 31, 1650097 (2016)

\bibitem{MK2}
E.O. Kahya et al, The European Physical Journal C 75, 43  (2015) 

\bibitem{MK3}
M. Khurshudyan, R. Myrzakulov, The Eur. Phys. J. C 77: 65 (2017)  

\bibitem{MK4}
M Khurshudyan, R Myrzakulov arXiv:1509.07357

\bibitem{MK5}
M. Khurshudyan et al, Eur. Phys. J. Plus 129: 119 (2014) 

\bibitem{Me11}
M. Khurshudyan, Symmetry, 8(11), 110 (2016)

\bibitem{MK6}
J. Sadeghi et al, Int J Theor Phys 55, 81 (2016)

\bibitem{BOSS}
T. Delubac et al.: Astron. Astrophys. 574, A59 (2015)

\bibitem{BOSS1}
E. G. M. Ferreira et al, Phys. Rev. D 95, 043520 (2017) 

\bibitem{DE1}
J. Yoo,	Int. J. Mod. Phys. D 21, 1230002 (2012)

\bibitem{SahniFin}
V. Sahni, A. Shafieloo, and A. Starobinsky, Astrophys. J. 793, L40 (2014).

\bibitem{PC}
Planck Collaboration, A$\&$A 594, A13 (2016)

\bibitem{U1}
M. Szydlowski et al, arXiv:1704.05364

\bibitem{U2}
K. Urbanowski, M. Szydlowski, AIP Conf. Proc. 1514, 143 (2013)

\bibitem{U3}
K. Urbanowski, Theor Math Phys 190: 458 (2017)

\bibitem{U4}
K. Urbanowski, PRL 1007, 209001 (2011)

\bibitem{U5}
A. Stachowski, Eur. Phys. J. C 77: 357 (2017) 

\bibitem{U5_1}
M. Szydlowski, A. Stachowski, Physics of Dark Universe, 15:96-104 (2017)

\bibitem{U5_2}
M. Szydlowski, Phys. Rev. D 91, 123538 (2015)

\bibitem{U5_3}
M Szydlowski et al, Journal of Physics: Conference Series 626, 012033  (2015)

\bibitem{U6}
K. Urbanowski, arXiv:1509.03830 

\end{thebibliography}
\end{document}